\documentclass[acmsmall]{acmart}
\AtBeginDocument{%
  }

\acmJournal{JACM}
\acmVolume{37}
\acmNumber{4}
\acmArticle{111}
\acmMonth{8}

\newcommand{\revision}[1]{{\color{black} #1}}

\usepackage{enumitem}

\begin{document}

\title{Understanding the Perceptions of Trigger Warning and Content Warning on Social Media Platforms in the U.S.}

\author{Xinyi Zhang}
\authornote{Authors contributed equally to this research.}
\affiliation{%
  \institution{Virginia Tech}
  \city{Blacksburg}
  \state{Virginia}
  \country{USA}
  }
\email{xinyizhang@vt.edu}

\author{Muskan Gupta}
\authornotemark[1]
\affiliation{%
  \institution{Virginia Tech}
  \city{Blacksburg}
  \state{Virginia}
  \country{USA}
  }
\email{95muskangupta@gmail.com}

\author{Emily Altland}
\authornotemark[1]
\affiliation{%
  \institution{Virginia Tech}
  \city{Blacksburg}
  \state{Virginia}
  \country{USA}
  }
\email{emilya99@vt.edu}

\author{Sang Won Lee}
\affiliation{%
  \institution{Virginia Tech}
  \city{Blacksburg}
  \state{Virginia}
  \country{USA}
  }
\email{sangwonlee@vt.edu}

\renewcommand{\shortauthors}{Zhang et al.}

\begin{abstract}
    The prevalence of distressing content on social media raises concerns about users' mental well-being, prompting the use of trigger warnings (TW) and content warnings (CW). 
However, inconsistent implementation of TW/CW across platforms and the lack of standardized practices confuse users regarding these warnings. 
To better understand how users experienced and utilized these warnings, we conducted a semi-structured interview study with 15 general social media users. 
\revision{
Our findings reveal challenges across three key stakeholders: viewers, who need to decide whether to engage with warning-labeled content; posters, who struggle with whether and how to apply TW/CW to the content; and platforms, whose design features shape the visibility and usability of warnings.
While users generally expressed positive attitudes toward warnings, their understanding of TW/CW usage was limited.
Based on these insights, we proposed a conceptual framework of the TW/CW mechanisms from multiple stakeholders' perspectives.
Lastly, we further reflected on our findings and discussed the opportunities for social media platforms to enhance users' TW/CW experiences, fostering a more trauma-informed social media environment.
}
\end{abstract}



\received{29 October 2024}
\received[revised]{12 March 2009}
\received[accepted]{5 June 2009}

\maketitle

\section{Introduction}

\begin{figure}[t]
  \centering
  \includegraphics[width=\textwidth]{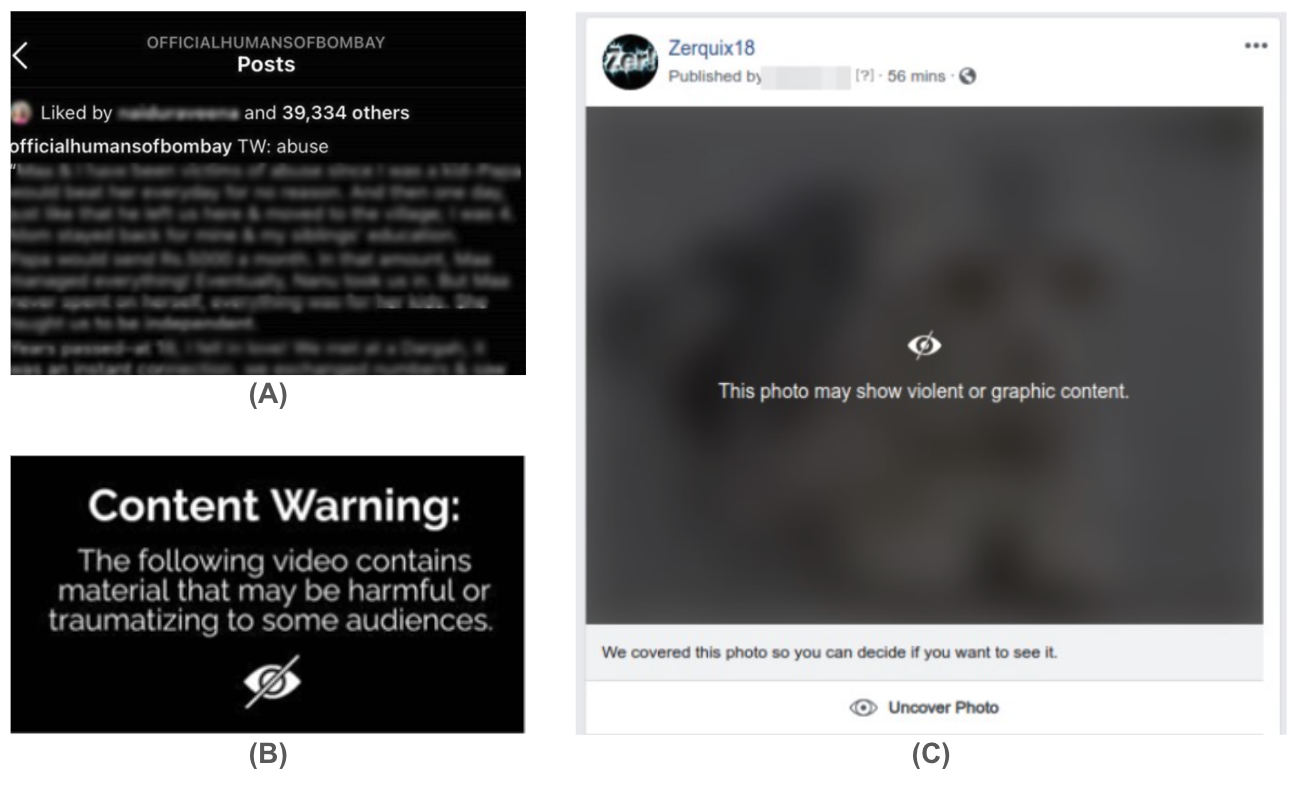}
  \caption{(A) TW for a text post on Instagram, (B) CW for a video on Facebook, (C) CW for an image on Facebook}
  \label{twcw}
\end{figure}

\revision{
Trigger warnings (TW) and content warnings (CW) are defined as statements presented as labels to help individuals avoid engaging with content that may evoke distressing emotions~\cite {bridgland2024meta}.
These warnings are widespread across multiple sectors, including health, education, media, arts, and literature~\cite{charles2022typology}.
The effective use of TW/CW relies on an understanding of trauma---the experiences and resulting aftermath of an extremely distressing event or series of events~\cite{chen2022trauma}.
Such events may include violence, abuse, neglect, loss, disaster, war, and other emotionally harmful experiences~\cite{huang2014samhsa}.
For individuals who have experienced trauma, exposure to related content can evoke emotional or physiological responses.
\revision{In addition, content warnings (CWs) are often applied to non-traumatic content that may still negatively impact or discomfort viewers. For example, content featuring strobe effects or rapid flashing lights may pose risks for individuals with photosensitive epilepsy~\cite{south2023exploratory}.
}

With the rise in the use of social media, especially with the pandemic~\cite{tankovska2021social}, distressing reactions can increasingly impact people’s experiences with technologies~\cite{chen2022trauma}. 
With different forms of content stimulating our senses, social media platforms can be risky for around 70\% of people who have experienced a distressing event~\cite{benjet2016epidemiology}. 
For example, individuals who have prior sexual assault experiences may avoid encountering content where others share their similar experiences on social media.
Exposure to distressing or triggering content on social media can negatively impact people's mental health \cite{zhao2020social, pettyjohn2022exploring}. 
To mitigate the potential harm caused by triggering content, social media users utilize trigger and content warnings (TW/CW) to warn those who may get triggered by sensitive content~\cite{charles2022typology}, including either themselves or others.
As illustrated in Figure~\ref{twcw}, TW/CW can be added into content in multiple ways, depending on the content modality (text, image, video). 
}
\label{Figures}

Given the recent emergence of TW/CW on social media, the information regarding their usage is relatively unknown. 
The diverse norms and modalities of warnings across different platforms suggest that \textit{``there may be a lack of clear and shared understanding among social media users on the usage of warnings.''} 
To foster safe social media interactions, we need detailed insights into the social media users’ thoughts and preferences around warnings~\cite{haimson2020trans}. 
Given the ubiquitous nature of such warnings on social media, it is essential to understand how users utilize TW/CW and what challenges they face when engaging with TW/CW.  


To better understand the challenges associated with warning usage and the intricacies of viewing content with such warnings on social media, we conducted a semi-structured interview study with 15 general social media users (not professional content creators).
In this study, we aimed to answer the following research questions:

\revision{
\begin{itemize}
    \item \textbf{RQ1:} How do \textbf{viewers} on social media decide whether to engage with content that includes trigger and content warnings (TW/CW)?
 (\ref{4.1})
    \item \textbf{RQ2:} How do \textbf{posters} on social media decide whether to include trigger and content warnings (TW/CW) when posting potentially triggering content?
  (\ref{4.2})
    \item \textbf{RQ3:} How can \textbf{social media platforms} enhance users’ experiences with trigger and content warnings (TW/CW)?
  (\ref{4.3})
\end{itemize}
}

While efforts have been made to address sensitive content through warnings, previous research has only focused on particular features (e.g., spoiler tags on Reddit, Facebook Memories~\cite{lustig2022designing}) or specific topics (e.g., LGBTQ+ experiences~\cite{haimson2020trans}, sexual violence~\cite{stratta2020automated}, and substance use recovery~\cite{phelan2022work}).
These studies provide valuable insights into the design of TW/CW systems within narrow contexts.
However, a gap remains in understanding how general users experience and engage with trigger and content warnings (TW/CW) across social media platforms in broader, real-world contexts.
Filling this knowledge gap is critical for the development of sociotechnical interventions that can effectively integrate TW/CW into social media environments, enabling users to navigate sensitive content on these platforms with greater autonomy and safety.
Through qualitative insights drawn from users' experiences with warnings, our study brings several key contributions to the understanding of the TW/CW system on social media:

\revision{
\begin{itemize}
    \item We examined the challenges that viewers faced when engaging with content labeled with TW/CW, which impacted their decision-making. 
    \item We investigated the challenges that posters faced when labeling content with TW/CW.
    \item We identified potential opportunities that social media platforms can implement to enhance users' (i.e., viewers, posters) experiences with TW/CW.
    \item Based on our findings, we proposed a conceptual framework that captured the TW/CW mechanisms from the perspectives of multiple stakeholders, including viewers, posters, and platforms.
    This framework can be used to inform the design of more effective warning systems.
\end{itemize}
}
\section{Review of Literature}

\subsection{Background On TW/CW Usage}

In academic literature, `trigger warning' (TW) and `content warning' (CW) are often used interchangeably, although some distinguish trigger warnings as a subset specifically targeting individuals who have experienced trauma or suffer from post-traumatic stress disorder (PTSD)~\cite{charles2022typology}. 
A recent systematic review analyzed 2,209 warnings across diverse sectors and categorized them into 14 distinct content warning types~\cite{charles2022typology}.

The literature presents contrasting views regarding the utility of warnings. 
On the one hand, advocates claim that warnings allow people to prepare themselves
, reduce negative reactions toward content~\cite{bridgland2019expecting}, and increase individual agency in making informed decisions about engaging with content ~\cite{charles2022typology}. 
On the other hand, critics suggest that warnings may lead to negative interpretations and encourage avoidance, potentially causing harm~\cite{bridgland2019expecting, bridgland2022meta}. 
Educational contexts have shown that warnings can increase anxiety and reinforce trauma's centrality in an individual's identity~\cite{charles2022typology}. 
However, the effectiveness of trigger warnings may be influenced by our shared cultural understanding, suggesting that their benefits or drawbacks can go beyond their initial clinical purpose~\cite{bridgland2022meta}.

While trigger warnings have been studied in clinical settings ~\cite{vlodder2023social, hyland2023qualitative}, their prevalence and impact among the general population in online spaces, such as social media platforms and online forums, remain underexplored, despite their widespread use in these digital environments~\cite{pasquetto2020tackling, george2020deciphering}.

To advance this discussion, it is necessary to shift away from the oversimplified discourse on whether content warnings have a positive or negative impact. 
Instead, a nuanced understanding of how content warnings affect different individuals in various contexts and when used for different reasons is required~\cite{charles2022typology}.

Therefore, our study employed a qualitative approach to gain a nuanced understanding of how users utilize TW/CW and engage with content featuring these warnings on social media through direct conversations with them.

\subsection{The Role of HCI in Social Media Warning Mechanisms}

\subsubsection{\textbf{Sensitive Triggering Content: A Gray-Area}}
Social media contains diverse sensitive content, ranging from controversial discussions (politics, religion, race, gender) to age-restricted material (nudity, pornography) and graphic imagery (violence, gore). 
Due to the diversity of personal experiences, content that may be acceptable to some individuals can also trigger traumatic memories in others.
Content around reflection systems is one such example that triggers social media users rather unexpectedly. 
For example, technology-mediated reflection (TMR) systems (e.g., Facebook’s Memories) curate specific past content for you to reflect on~\cite{lustig2022designing} may evoke users' bittersweet emotions, even when the memories positive (e.g., wedding photos) but associated with grief (e.g., divorce, death, severe illness).

Dealing with explicit content through moderation (e.g., removing or hiding posts) may seem straightforward, but certain sensitive content falls into a gray area concerning social media site policies and community norms even though sometimes they do not violate site policies~\cite{haimson2021disproportionate}. 
For example, content related to the transition experiences of transgender individuals can be considered gray-area material. 
While crucial for LGBTQ+ visibility, education, and activism, it can inadvertently clash with nudity guidelines or be subject to reporting and blocking due to discrimination against personal identities~\cite{haimson2020trans}. 
Moreover, sensitive topics like self-harm and suicide present a dilemma between censorship to protect vulnerable viewers and the need to raise awareness for prevention~\cite {keelytedx}. 

Another form of gray area content is tied to past trauma, which is context-dependent and challenging to predict. 
For instance, individuals recovering from substance use disorder may encounter images of substance use on social media, unexpectedly triggering their recovery progress~\cite{phelan2022work}. 
This demonstrates that algorithmic knowledge can inadvertently harm users, as seen with targeted weight-loss advertisements negatively affecting people with histories of disordered eating~\cite{gak2022distressing}.
Additionally, while TW/CW is often associated with sensitive content \cite{hyland2023qualitative}, moderating such content and navigating triggers online is challenging. 
\revision{
Research by South et al. highlights this complexity by showing that generic warnings may not always be effective in the context of photosensitive epilepsy.
They found that warnings specifying the scenic and temporal aspects of a video were more helpful than general alerts~\cite{south2023exploratory}. }
Therefore, a comprehensive examination of factors influencing warning effectiveness is essential for supporting viewers' decision-making.
Our research focuses on a deeper understanding of user risks and vulnerabilities when engaging with potentially triggering content, regardless of the presence or absence of TW/CW labels.

\subsubsection{\textbf{Personalized Filtering}}
Trauma-informed approaches are generally seen as beneficial for all people, regardless of whether they are trauma survivors \cite{huang2014samhsa}. 
Trauma-informed computing recognizes that digital technologies can both cause and exacerbate trauma and seeks out ways to avoid technology-related trauma and retraumatization \cite{chen2022trauma}. Chen et al.'s trauma-informed computing framework suggests that engineers and designers could follow the principles of collaboration and enablement to incorporate people’s conscious choices into their information feeds. 
They also warn that content filtering can cause \textit{"filter bubbles"}, limiting the content people are exposed to, often without their awareness \cite{chen2022trauma}.
On the contrary, Randazzo et al.\cite{Randazzo_Ammari_2023} argue that while filter bubbles may polarize, filtering algorithms can be beneficial for trauma survivors by empowering them to challenge societal filters imposed by institutions. 
For instance, people in substance use recovery may encounter subtle triggers in digital environments that can accumulate and jeopardize their progress if encountered at the wrong time \cite{phelan2022work}. To enhance user safety implementing advanced content filters \cite{Randazzo_Ammari_2023, phelan2022work} and semi-automated trigger warnings \cite{phelan2022work} can be valuable. 
Despite the presence of basic content filters on social media, Phelan et al. \cite{phelan2022work} observed that participants frequently encountered triggering content, highlighting the limitations of these tools.

Our paper aims to enhance the design of content filtering and other warning-related features by providing insights into the challenges of TW/CW utilization and the dynamics of users' interactions (e.g., viewing or skipping) with warned content on social media platforms.

\subsection{Social Computing Systems to Keep Users from Triggering Content}

Some social media platforms already have tools to filter out certain users easily. 
\revision{For example, Twitter allows users to use a blocklist, a list of preemptively blocked accounts from interacting with a subscribe, to avoid harassment/triggering content~\cite{jhaver2018online}. }
However, these features do not allow people to hide only certain posts from a user - it is either blocking the entire content from the specific user or nothing.
To address the need for personalized content management, systems like \textit{SquadBox} utilize \textit{friend-sourced moderation}---blocking harassment emails from selected users~\cite{mahar2018squadbox}---as a possible solution to combat challenges. 
However, this worked effectively for direct contact management, friend-sourced moderation struggles to handle an overwhelming volume of content on social media platforms.

Warnings offer an alternative for sensitive content without removal, aligning the site guidelines to protect users who might be triggered. 
For example, the social media platform, \textit{TransTime}, designed for the transgender community, employed \textit{Content Warning Tags}--a hashtag-like feature that allowed users to filter posts based on the content tags~\cite{haimson2020trans}. 
Crowdsourced labeling of sensitive content has also shown promise.
\textit{Shinigami Eyes}, a browser extension, uses community-feedback-driven to color-code transphobic and trans-friendly social media users and content~\cite{saetrashinigami, shinigamieyes, bowker2022reducing}. 
Similarly, \textit{DeText}, uses keyword recognition and sentiment analysis to auto-generate content warnings related to sexual violence~\cite{stratta2020automated}. 
However, these tools are topic-specific, focusing on trans-issues, and sexual violence, respectively.  
Another website, \textit{DoesTheDogDie.com}, allows users to pre-review movies, TV shows, and video games across over 100 trigger categories~\cite{doesthedogdie}, ranging from violence and abuse to amputation and broken bones, even a dragon dies or shaky cam. 
It also includes hotlines as a resource for topics like self-harm~\cite{doesthedogdie}. 
While highly effective for entertainment content, there is no equivalent for social media platforms.

In the existing literature, TW/CW has been explored within the context of social computing systems, often as a specific feature or in relation to particular cases or topics. 
However, there remains a gap in understanding the border use and impact of TW/CW among general social media users.

\subsection{Knowledge Gap}
In conclusion, the discourse surrounding TW/CW usage on social media highlights the intricate challenges associated with navigating sensitive content, with or without warnings. 
While previous research has explored TW/CW feature designs for specific case(s), examined content filtering, and underscored the necessity for nuanced approaches, gaps lay behind the existing solutions. 
Even though various topics (e.g., transphobia, photosensitivity, substance use disorder, and histories of disordered eating), have been examined individually, the common challenges that transcend these specific contexts require further investigation. 
Our study fills these gaps by offering holistic insights into refining TW/CW features that facilitate content viewing and assisting content creators in applying warnings effectively.

\section{Study Design} 

Because trauma is complicated and we know comparatively less about trauma-informed interactions on social media~\cite{scott2023trauma}, we designed our research to be qualitative, which \textit{``is appropriate when the purpose of the research is to unravel complicated relationships''} \cite{rubin2011qualitative}.
We conducted 15 semi-structured interviews with general social media users to understand their experiences associated with trigger warning (TW) and content warning (CW) on social media. All the interviews were conducted after getting approval from our university's Institutional Review Board.


\revision{

\subsection{Participant Recruitment and Eligibility}
We recruited participants through our university's graduate student mass email listserv and `Computer Science Discord' server with a short study description, mentioning the eligibility criteria (specified in the following section) and a survey link (see appendix \ref{sec:survey}).
We also distributed the same study materials (i.e., recruitment advertisements and survey questionnaires) through our personal Facebook and Twitter accounts.
Additionally, we advertised the study on Reddit via the `r/mentalhealth' subreddit.

Participants were eligible for this study if they were 18 years or older, English speaking, currently US citizens or legal residents, and regular social media users who have added or seen TW/CW on posts. 
The survey also included an informed consent form, requiring potential participants to sign before participating in the interview study. 
From the initial 181 responses collected, we filtered out fraudulent submissions.
Using a purposive sampling approach, we selected and conducted 15 participants for online interviews, considering their diverse opinions and backgrounds, race and ethnicity, gender, and social media use and experiences.

All participants lived in the United States at the time of the interviews. Their ages ranged from 21 to 33 (M = 25.46, SD = 3.34). Seven
participants identified as male, seven as female, and one as non-binary. 
The interviewee details are presented in Table \ref{table:1}.
We also gathered information about their social media platform usage, posting, viewing frequency, and their experiences with TW/CW. 
All of our participants were regular social media users rather than professional content creators or those who were traumatized by something and avoided TW/CW content seriously.
They all had encountered TW/CW in some form, and eight of them had posted content with TW/CW at some point.
We attached the participants' detailed background information in the Appendix~\ref{background}.
We did not inquire about their mental health diagnoses. 
However, they had the option to discuss this during the interview if they were willing to.

}

\subsection{Interview Protocol}

\revision{
Due to the sensitive nature of the interview, we paid utmost attention to make sure that interviewees were protected from potentially traumatizing experiences. 

\textbf{Consent Process Before the Interviews}
Before starting the interview, we warned that the interview may involve discussion around triggering topics in advance prior to getting consent. 
In addition, we asked them if there were particular topics that they did not want to discuss.
Then, we informed participants that their participation was completely voluntary and they were free to skip any questions and withdraw from the study anytime without any penalties.
Then, we interviewed all the participants via Zoom, and each one lasted in the range of 50 minutes to 1.5 hours.

\textbf{Initial Interview Protocol with the First Six Participants.}
In the first six interviews, we prepared a set of questions that guided our semi-structured interview, which contained four aspects: (1) participants' typical social media use and how it differed by platform;
(2) participants' use of social media and trigger/content warnings as a viewer/consumer of content;
(3) participants' use of social media and trigger/content warnings as a poster/producer of content;
(4) participants' desirable user experience when it comes to how platforms should handle trigger and content warnings on social media.
\revision{The set of guiding questions is available in Appendix~\ref{sec: initial_interview}.}
}

\revision{
\textbf{Updated Interview Protocol with the Rest Nine Participants.}
We updated the interview protocol after conducting interviews with the first six participants based on the interviewers’ reflections and emerging themes from the initial coding. Most importantly, we revised the questions to better highlight the dual perspectives of users, as both content posters and viewers. Initially, our study focused on the perceived differences between trigger warnings (TW) and content warnings (CW). However, after the first phase of interviews, we did not find any meaningful distinction between TW and CW themselves. Instead, what emerged was a nuanced difference in how participants perceived TW/CW content depending on whether they were viewing it or posting it. In that regard, we updated our interview questions to capture the potential tension they may have as a poster and viewer more explicitly.  For example, we added a question such as ``What topics would you feel hesitant about adding TW/CW to in your posts, and why?" 
The updated interview protocol is available in Appendix~\ref{sec: interview}, and we also used bold text to indicate when the warning examples should be shown to the participants. 

In addition, we made some changes that will help the interview go more effectively. 
First, we developed a slide deck featuring various examples of how TW/CWs are displayed across existing social media platforms (see Appendix~\ref{slideshow}). 
Interviewers could use this visual aid to prompt discussion about the perceived effectiveness of different formats and participants' preferences as posters. These examples were collected from a range of platforms and organized by modality (text, image, video). Second, the interview questions were streamlined and condensed while retaining the core themes, making the interview more conversational and participant-centered.
}


\vspace{10pt}

\begin{table} [h]
\setlength{\tabcolsep}{8pt}
\centering
\begin{tabular}{cc p{7.5cm}p{2.5cm}}
\toprule
\textbf{PID} & \textbf{Age} & \textbf{Race or Ethnicity} & \textbf{Gender} \\ \midrule
P1 & 33 & White or Caucasian & Non-binary\\ 
P2 & 27 & Asian or Pacific Islander & Man/Male\\ 
P3 & 27 & South Asian & Woman/Female\\ 
P4 & 26 & Hispanic or Latino & Woman/Female\\
P5 & 28 & White or Caucasian, North Africa and Middle East & Man/Male\\
P6 & 30 & Black or African American & Woman/Female\\
P7 & 28 & Asian or Pacific Islander & Woman/Female\\
P8 & 23 & White or Caucasian & Man/Male\\
P9 & 25 & Asian or Pacific Islander & Man/Male\\
P10 & 21 & White or Caucasian & Woman/Female\\
P11 & 21 & Hispanic or Latino & Woman/Female\\
P12 & 22 & Asian or Pacific Islander & Man/Male\\
P13 & 25 & Asian or Pacific Islander & Man/Male\\
P14 & 22 & Black or African American & Man/Male\\
P15 & 24 & Asian or Pacific Islander & Woman/Female\\
\bottomrule
\end{tabular}%
\caption{Demographics of the Participants}
\label{table:1}
\end{table}

\subsection{Data Analysis}

Once data was collected, we conducted a thematic analysis~\cite{braun2006using}.
We adopted an inductive thematic analysis in two phases: the first phase analyzed the first six interviews (P1-P6), and the second phase covered the remaining nine (P7-P15).
This approach helped us derive insights from the collected interview data. 

We transcribed the interview recordings using Sonix.ai and manually corrected the transcriptions. 
Then, we open-coded a couple of transcriptions at the sentence level.
Once we generated the initial codebook, we utilized axial coding to explore the underlying themes and examine the potential relationships between these codes. 
As we analyzed the first two interviews, we identified gaps in our understanding, prompting adjustments to both the interview questions and participant recruitment to better address these gaps. 

Initially, our study focused on the differences between TW and CW.
This highlighted the complexity of triggers and the nuances found in TW/CW perception.
Therefore, we shifted our goal to explore the participants' decision-making processes and the factors that influence the effectiveness of TW/CW. 
In the end, we conducted the study and analysis in two iterative cycles.

After the first round of analysis, we grounded the codes into six initial themes:
\begin{enumerate}
    \item Overall perceptions of warnings on social media
    \item Topics covered by TW/CW
    \item Perceptions on adding a warning on sensitive content
    \item Perceptions around viewing a warning over sensitive content
    \item Factors that make a TW/CW effective
    \item Design recommendations surrounding warnings
\end{enumerate}

Then, we organized these six themes into a hierarchical tree structure, merging repetitive sections or paragraphs to develop the final three themes:

\revision{
\begin{enumerate}
    \item Factors influencing the viewer's decision-making process when interacting with content containing TW/CW.
    \item Factors influencing the poster's decision-making process when adding a warning before posting triggering content.
    \item Challenges caused by social media platforms and users' design needs for platforms.
\end{enumerate}
}


\section{Results}

Based on thematic analysis, we organized our findings from the perspectives of three stakeholders, including viewers, posters, and social media platforms. 
 
\subsection{RQ1: How Do Viewers on Social Media Decide Whether to Engage with Content that Includes Trigger and Content Warnings (TW/CW)?}
\label{4.1}
When a TW/CW is presented, the viewers have to decide whether to engage with the content or avoid it. 
This decision-making process, as we discovered in our data, is shaped by multiple factors in two general types, including the external factors relevant to how TW/CW is presented (e.g., specificity, modality) and a viewer's internal factors  (e.g., trust and familiarity with the content posters, personal connections, and relevance of the topic, tolerance, and current mental state). We present the details below. 

\subsubsection{\textbf{Accompanying Information Affects Viewers' Engagement with TW/CW}}
We identified several characteristics of the warnings that affect viewers' decisions on whether to engage with the content or not. 
The first is warning specificity. Specificity refers to the amount of details in the warning's description, including the content and the context. 
Participants demonstrated conflicting needs for the specificity of the warnings. 
Some participants preferred to have more general warnings with less specificity, while others preferred less general warnings with more details. 

\textbf{Including Specific Topics in TW/CW Reduces the Psychological Burden.}
\label{specificity of warning}
The specificity of warnings was highlighted as particularly important for informed decision-making.
During our interviews, participants' decisions to engage with or avoid content marked with TW/CW were strongly influenced by the specificity of the warnings. 
For example, most participants (12/15) indicated that insufficient information, which often comes with vague warnings, undermines viewers' autonomy and their ability to make informed decisions. 
For example, simply listing TW/CW without telling what the topic is about, viewers cannot accurately assess whether the content aligns with their personal triggers or experiences.
Precise warnings, on the other hand, were perceived to be effective, as knowing the exact nature of the post content helps viewers mentally prepare or decide to avoid potential distress proactively, mitigating negative emotional impacts.
For example, P14 illustrated this points, 
\begin{quote}
    \textit{``I think having a non-general warning or a specific warning is better because just having a general warning is like we don't know what's going on there, so we don't know if it's something we may or may not have experienced. I think having an experience of that might bring back the trauma.''} (P14)
\end{quote}
In this example, P14 emphasized that the lack of specificity in the warnings increased the uncertainty of the anticipated consequences if engaging with the warnings, which may even recall traumatized experiences. 
However, even when users specify content types, the terminology can be unclear, particularly when abbreviations like those used by P11 (e.g., SA for sexual assault, DV for domestic violence, ED for eating disorders) are employed.

Additionally, the absence of specific topics and severity information in automatically added warnings can also lead to a false sense of security (P3). 
As an example, P10 expressed concerns that overly vague warnings (e.g., ``Sensitive Content'' on Figure~\ref{slide1}-E and Figure~\ref{slide3}-D in Appendix) could either lead viewers to view content without realizing that it can be triggering or keep them from even when it is something that they can tolerate.


\textbf{TW/CW Itself Can Be Triggering If ``Too Specific.''}
However, some participants (3/15) acknowledged that overly detailed warnings could be problematic because the warning itself may potentially trigger the viewers when providing too many specifics in a warning. 
For example, P04 mentioned as follows when discussing an example given in slides:

\begin{quote}
    \textit{``Like, for example, the middle one (Figure \ref{slide1}-B), the one of the book is like `it's a true story focusing on sadistic torture and abuse'. Like if you go too specific with your trigger warning, you're going to trigger somebody.''}(P04)
\end{quote}
For example, a participant expressed discomfort when viewing a slide labeled ``TW: Gangrape (Figure \ref{slide1}-C)'', finding the wording too explicit. 
P13 said that the word gangrape is \textit{``too specific''} and \textit{``reading the word itself might be a trigger for someone''} (P13). 
Therefore, determining how specific a trigger or content warning (TW/CW) should be involves balancing two risks: the risk of triggering individuals if the warning is too vague and the risk of exceeding expectations or causing distress if it is too explicit.
For example, labeling a post as ``sexual abuse, graphic content'' can be a more moderate option than stating it ``gangrape''.

Similarly, four participants appreciated complete obscurity with no identifiable information visible, along with descriptors to enhance specificity, as they \textit{"don't want to have imagination sometimes about it"} (P7).

The result highlights a nuanced tension; while many emphasized the importance of detailed warnings for informed decision-making, others cautioned that excessive specificity could be triggering in itself. 
Striking a balance between providing enough context and avoiding overly explicit details is essential to achieving the intended protective function of TW/CW. 



\textbf{TW/CW Can Elicit Curiosity.}
If a warning is too general or the content is only partially obscured—especially in visual formats—it can provoke curiosity and lead viewers to engage with the content.
For instance, P12 found themselves more intrigued by what might be concealed behind the warning, whether it is sensitive images or topics, and this curiosity led them to watch the content.
    \begin{quote}
        \textit{``It makes me more curious about what's underlying like behind the blurred picture. If it's like skin color, maybe it's a nude; if it's a red color, maybe it's blood.''} (P12)
    \end{quote}
\noindent When viewers are presented with partial information, such as a blurred image with distinguishable shapes or color tones, they may feel drawn to explore further out of curiosity, thus risking exposure to harmful content.
The vague or contextless warnings are not effective and could encourage viewers to explore further.
This increases the risk of encountering triggering content and defeats the purpose of having a warning. 
Viewers can regret it later, get affected more than they thought, or feel betrayed by the author. 

A few participants (3/15) mentioned that they prefer warnings that completely obscure the image, with no visibility of the author, comments, or colors of the photo, to avoid evoking viewers' curiosity.
For example, P7 claimed that: \textit{``it will be helpful that we cannot see who posts it, because some people there may be curious or just try to find it''}.
These findings highlight a tension between the need to protect viewers from potentially distressing content and the risk that partial or vague warnings may unintentionally provoke curiosity, undermining their intended protective function.

Relatedly, some participants perceived that adding warnings may encourage users' engagement. 
Some participants mentioned that some spam bots use TW/CW as a strategy to grab attention, particularly on platforms such as Tumblr and Reddit. 
As P1 explained,
\begin{quote}
    \textit{``[...] if you follow or pay attention to a specific tag, then it'll pop up for you. So spam bots will use the specific tags so that you see their content.''} (P1)
\end{quote}
\noindent These findings suggest that while TW/CW are intended to protect, their design can backfire—triggering curiosity, encouraging unwanted engagement, and ultimately increasing the likelihood of exposure to distressing content.

\textbf{Content Modality Influences Viewers' Decisions.} 
The modality of a social media post also plays an important role in reviewers making informed decisions. 
Participants often made decisions to view or skip content with TW/CW based on the modality of the content. 
In our study, participants generally found it easier to engage with text-based content that included TW/CW, reporting fewer issues processing such material. 
For example, P4 shared her experience:
    \begin{quote}
        \textit{``I think it really depends on the format. I am more likely to read something with a trigger than I am to see something with a trigger warning. Especially because reading, I guess you can stop at any time, and watching is a little bit harder because you can close your eyes and you still have audio, right?''} (P4)
    
    
    \end{quote}
In summary, participants’ responses indicated that content modality influenced their decision to engage with material featuring TW/CW. Text-based content with warnings was generally perceived as less distressing, while participants were more cautious about engaging with videos and images. However, this perception may lead to the unintended consequence of overlooking content warnings for textual content, which can still be deeply distressing or triggering for some viewers.

\subsubsection{\textbf{Challenges in Warning Discoverability for Viewer Decision-Making}}
Participants highlighted the importance of visibility and placement in determining the effectiveness of TW/CW.
Warning discoverability refers to the extent to which warnings are discoverable to viewers within the posts.
Prominent and easily distinguishable warnings enable viewers to make informed decisions about whether to engage with content.
For example, several participants (5/15) noted that font size (P12) and visual cues such as emojis (P13) helped in quickly capturing viewers' attention (e.g., Figure~\ref{twcw}-B).
These features help visually interrupt the scrolling experience, making the warnings stand out before viewers engage with the warning-marked content.
As P12 explained, visual warnings \textit{``alert me more than text''} are therefore more effective in preventing accidental exposure to sensitive content.

Another challenge we identified is that TW/CW added in text can be easily overlooked--especially on visually driven platforms like Instagram, TikTok, or YouTube--where viewers tend to focus on images or videos before noticing any accompanying text. Three participants noted that they found the poster-added warning in the caption/description not helpful in deciding, since the viewer's attention was more towards the visual content than the warning in the caption text.
For example, in the Instagram example shared in the slide (Figure \ref{twcw}-A), P10 critiqued the ineffectiveness of how the warnings were just in the caption:
\begin{quote}
    \textit{``I can already see the immediate issue with Instagram because it puts the picture first and the text is very small. You're going to read the image before you even have time to look at the description.''} (P10)
\end{quote}
\noindent 
P14 further pointed out that when visual content was reposted on Instagram stories, the caption was often hidden by default, making the warning even less visible to viewers.
We discuss relevant challenges that posters will have in adding TW/CW in primarily visual platforms in~\ref{poster_logistics}.


\subsubsection{\textbf{Viewers' Personal Interests/Background Affects Their Engagement with TW/CW}}

Participants' decisions to engage with or avoid content labeled with TW/CW were influenced by a range of viewers' personal factors, including their tolerance to and interest in potentially distressing content, the content posters or topics, and their current mental state.

\textbf{Viewers' Tolerance Varies by Topic.}
\label{viewer tolerance}
Participants' decision to view the content behind the warning was influenced by their personal ability to tolerate sensitive material as well. 
Participants varied in their responses, with some demonstrating high tolerance for distressing material while others were more vulnerable to emotional impact.
Four participants indicated that they tended to view content with TW/CW without getting easily triggered, as they had a high tolerance for such content and rarely got triggered. 
As a result, they are inclined to watch content with TW/CW so that they can handle it without adverse emotional reactions.
For example, P12 highlighted his experience with high tolerance.
    \begin{quote}
        \textit{``It was a generic warning, like trigger warning, `Do you really want to watch?' And I press `Yes' because I think I have a high tolerance for those kinds of things. I don't get triggered that easily.''} (P12)
    \end{quote}
\noindent The comment reflects a degree of emotional preparedness and confidence in their mental boundaries. 
This high tolerance also meant that they were less likely to suffer from seeing TW/CW, and were more resilient to sensitive content.

Whereas certain potentially triggering topics with TW/CW, to which the participants had a direct and sensitive connection, were avoided at all costs. For example, a participant who has struggled with suicidal tendencies chooses to steer clear of content related to suicide as part of their ongoing recovery journey (P7). However, they were more open to viewing other sensitive content.

    
    
    
    
Other participants decided to skip viewing the sensitive content, or at least save it for later, when they had a low tolerance for that material, and it could have a detrimental effect on their mental well-being. 
For instance, P2 made a conscious choice to skip content with TW/CW because they believed that consuming such material would have an adverse effect on their mental health. 
    \begin{quote}
        \textit{``I mostly skip. I don't like to. I mostly skip all that content because I feel like consuming that content has a negative effect on me.''} (P2)
    \end{quote}

In summary, participants' choices regarding whether to engage with or skip content with TW/CW were influenced by their emotional tolerance. Those with higher tolerance felt more resilient to triggering content, whereas those with lower tolerance were more cautious and opted to skip viewing such content to protect their mental well-being.

\textbf{Relevant Topics Encourage Engagement.}
\label{relevent to topics}
Another factor was the degree of personal connection to the topic.
In certain instances, participants opted to view content flagged with warnings, driven by their desire to stay informed about relevant topics given in the descriptor of the warning labels. 
Whether it was a local event, like an incident in their hometown (P2), or a campus-related matter, such as a sexual assault case (P15), they chose to engage with such content to understand the situation better and support relevant causes. 
One participant (P7) also watched sensitive content to gain knowledge on a specific issue, like child abuse, demonstrating their commitment to learning about important societal matters. 
Furthermore, individuals like P12 engaged with content marked with TW/CW as a means to stay aware of global events, even when the subject matter is distressing or unpleasant.

    

In contrast, some participants ignored TW/CW altogether if the topic was not personally relevant. 
This reflects a perception that trigger warnings are most valuable to individuals who have interests in, if not experienced, the specific type of trauma referenced. As P14 shared:
\begin{quote}
    \textit{``The trigger warnings seem like they're more for people who might have experienced this. So when it comes to warnings about certain things that I might not have experienced, then I'm more inclined to ignore the warning.''} (P14)
\end{quote}

Overall, participants' choices to engage with content featuring TW/CW partly depended on their personal connection to the topic. If they wanted to learn more about the topic, they would ignore the warning even when it could affect them. Certain topics were avoided at all costs due to personal history with them. In contrast, some participants tended to skip or ignore warnings when the content did not directly apply to their experiences. 

\textbf{Familiarity with the Posters Facilitates Informed Decisions.}
\label{connection with poster}
One influential factor was the identity of the author or poster. 
Participants explained that when the warning came from someone they knew, such as friends or family members, they were better able to anticipate the type of content behind the warning.
This familiarity helped them make more informed decisions about whether to engage with or skip the post. 
For instance, P4 shared that they had friends who regularly posted surgical images from their medical school experiences. 
When they encountered a blurred image with a warning from one of these friends, they anticipated graphic surgical content and chose to skip it based on their experience with the poster's previous social media posts.
    \begin{quote}
        \textit{``I have maybe three or four friends back home who went to medical school and, for some reason, like to post pictures of people being opened up in the surgical room. So I know that if it's one of them and the images blurred out and it's a content warning, I'm like, oh, I'm about to see some intestines out of somebody, and I hate that. So I skipped immediately.''} (P4)
    \end{quote}
\noindent Warnings attached to content from unknown or random posters evoked more hesitation and uncertainty than the ones that they know of.
This distinction highlights the role of trust and familiarity with authors in making a decision to view or skip content with warnings.

\textbf{A Stable Mental State is Necessary to View TW/CW Posts.}
\label{viewer mental state}
Some participants revealed that their decision to engage with potentially triggering content depended on their current mental state and overall mood. 
When feeling stable and open to learning, they might choose to watch such content for educational purposes. 
They consider factors like whether they have taken their antidepressant medication or not. 
However, during periods of vulnerability or low mood, they opt to avoid it.
Moreover, it also depends on their brain power to process and unpack warnings to determine whether the content will affect them or not.
For example, P15 explained:
\begin{quote}
    \textit{``It's also depending on the headspace and the mood. Like, if I'm already sad, like, what is the point of reading more? Consuming more negative information. And also, if I'm just really tired and just on social media to like to kill time before I go to bed, I might not have the power, the brain power to actually process and unpack the trigger warnings.''} (P15)
\end{quote}
\noindent This highlights that decisions are not static but instead can change dynamically depending on viewers' internal conditions at a given moment. 
This dependence can be relevant to the previous point that people may view such posts with the risk of being distressed when they are interested in the topics; their mental state can affect the decision to view the content that they are interested in or not, even when it can potentially distress them.

\revision{

\subsection{RQ2: How Do Posters on Social Media Decide Whether to Include Trigger and Content Warnings (TW/CW) When Posting Potentially Triggering Content?}
\label{4.2}

In our study, the majority of participants (12/15) acknowledged that TW/CW is an essential approach to protect viewers from unexpected exposure to sensitive or distressing content.
These warnings enable viewers to make informed decisions about whether to engage with such content, allowing them to protect their mental health.
Although a small fraction of participants found it relatively straightforward to determine whether to add TW/CW or not, as an author of social media posts, the majority faced challenges involved in using TW/CW: identifying which topics need warnings and navigating the balance between providing warnings and maintaining user engagement added complexity to the endeavor.

\subsubsection{\textbf{Challenges in Identifying What Topic Needs Warning}}
\label{warning types}

The participants emphasized a main challenge they faced when posting content: the lack of a consistent perspective or standard regarding what types of content should be labeled with warnings.  
Due to the impossibility of knowing all potential triggers, they found it difficult to determine which warnings were necessary, leading to inconsistent use of warnings or hesitation to add them at all.
As P3 explained,
\begin{quote}
    \textit{``What is a trigger for me might not necessarily be a trigger for you.''} (P3)
\end{quote}
\noindent This quote further explained that the diverse and subjective nature of triggering topics contributed to the complexity of determining when a topic warrants a warning, as individuals must navigate different interpretations of what may be considered triggering. 

As shown in Table~\ref{table:2}, the participants identified certain topics, i.e., Types 1–7, that they generally agreed should be labeled with warnings. These topics---such as violence, abuse, and self-harm---align with the content warning categories synthesized by Charles et al.~\cite{charles2022typology}. For content related to these topics, the expectation to include warnings was widely accepted.
However, for some types, i.e., Types 8-11, only a subset of the participants mentioned that they needed TW/CW for the topics. These topics vary depending on a user's background, personal experiences, and community, making it more challenging to generalize warnings for potentially sensitive content.
As P11 explained, 
\begin{quote}
    \textit{``There's not really a clear answer or clear guidelines to what a trigger is. You know, like, a paper cut might be really triggering for someone, but it might take like a really gory picture to be triggering for someone else.''} (P11)
\end{quote}
\noindent This illustrates that even when discussing the same topics, individuals may differ in their perceptions---some think a TW/CW is necessary, while others do not.
}

\vspace{10pt}
\begin{table} [h]
\setlength{\tabcolsep}{8pt}
\centering
\begin{tabular}{c p{4.6cm} p{6.8cm}}
\toprule
\textbf{ No.} & \textbf{Warning Types} & \textbf{Sub-Types} \\ \midrule
1 & Violence, Abuse, and Graphic Content & animal abuse, sexual violence, rape, sexual abuse, child abuse, mental abuse, domestic violence, domestic abuse, murder, instigation of violence, police brutality, gun violence \\ \midrule
2 & Suicide and Self-Harm & \\ \midrule
3 & Marginalization-Based Violence and Hate & people of color, children, LGBTQ+ people (homophobia, transphobia), religious minorities (Islamophobia), hate crimes, ableism, sexism, racism, harassment of women \\ \midrule
4 & Photosensitive Epilepsy Triggers & \\ \midrule
5 & Grief and Trauma Incidents & illness, death \\ \midrule
6 & Political Content & white supremacy, eugenics, and political beliefs\\ \midrule
7 & Immigration and Deportation & \\ \midrule
8 & Phobias & holes, needles, insects, birds \\ \midrule
9 & Natural Calamities and War & car crashes, plane crashes, war-related content \\ \midrule
10 & Mass Shooting & gun violence \\ \midrule
11 & Audio Triggers & audio includes trigger content \\ 

\bottomrule
\end{tabular}%
\caption{Types of Trigger Warning (TW) and Content Warning (CW) Identified from the Participants }
\label{table:2}
\end{table}

\revision{
\subsubsection{\textbf{Logistical Challenges for Posters Adding a Warning}}
\label{poster_logistics}
In our study, participants found it challenging to include warnings when creating a post due to the absence of a consistent standard for TW/CW across platforms. 
Different platforms employ various methods to display warnings for the same triggering content, which creates unique challenges.
These challenges may cause posters frustration.
This inconsistency not only increases the cognitive and logistical burden on content posters but also reduces overall effectiveness and motivation to add warnings.

\textbf{Limited Built-in Features for Adding TW/CW} 
The extent to which social media support adding TW/CW varied per platform. 
Only a few platforms provided a built-in feature to flag their visual content with a warning. 
The most common approach (12/15) was blurring images where users can voluntarily flag their social media as ``sensitive content,'' and images or videos were initially blurred until users \textit{``chose to interact with them''} (P8). 

Text fields (11/15), such as captions, titles, or descriptions, were commonly used to provide warnings, although they took different approaches. For example, P1 placed hashtags as warnings below an image, while P10 put the label upfront before the content. Some participants tried to emphasize the warning by putting it in uppercase letters (P5) or using abbreviations (e.g., SA for sexual assault) for specific topics (P11).

We also found that users re-purpose existing methods that are not necessarily designed for TW/CW to warn viewers. 
One example was using the spoiler feature (Figure \ref{slide2}-D); some people blocked the entire content of the text, while some blocked parts of it. 
P10 mentioned that the spoiler feature on Reddit or Discord is commonly used to conceal triggering information. They also noted that \textit{``if you mark the link in spoiler tags on Discord, it will blur the thumbnail''}. 
The label spoiler can miscommunicate the intention of why an image is blurred or some text is redacted, but at least it can keep the content from being exposed and allow users to learn why these were blurred from the context, most of the time with TW/CW, as in See Figure~\ref{slide2}-A and D. 
This range of preferences and hacked solutions highlights the challenges in determining the most appropriate warning method in different contexts.

\textbf{Custom TW/CW When for Visual Content} 
For visual or multimedia content, the participants have noticed custom solutions to effectively display TW/CW as part of the content
Several participants (8/15) also employed the ``first-slide'' method, where they dedicate the first slide of an image-based post that involves a slideshow of multiple images for the warning before the actual sensitive content (P10). 
Additionally, spoken warnings were not uncommon in videos, especially on platforms like Instagram and TikTok, where creators would verbally announce a trigger warning before discussing sensitive topics (P4). 
These diverse methods reflect the various strategies participants used to ensure that warnings were added to multimedia content.
However, the wide range of approaches contributes to the complexity of participants' ability to consistently apply warnings across different types of content and platforms.

}

\revision{

\subsubsection{\textbf{Tension of TW/CW with Engagement}}
The participants had varying opinions on how adding TW/CW affected engagement on a post, presenting challenges in deciding whether to include warnings. 
Most believed that adding a warning would reduce views as platforms decreased content visibility, while a few participants felt that warnings could increase engagement, as some of their viewers appreciated the ability to make informed decisions before interacting with the content.

In our study, some participants noted that they might avoid adding TW/CW warnings due to concerns that it could limit the reach of their content, as platforms often deprioritize sensitive content. 
For instance, P2 suspected that Instagram's algorithm may rank sensitive content lower, resulting in reduced visibility, even for users they follow who share such content:

\begin{quote}
    \textit{``I don't think I've seen trigger warnings in the past two months on Instagram. I mean, I do follow some pages that might post like triggering stuff, but I think they're algorithmically trying to reduce reach.''} (P2)
\end{quote}

\noindent Looking at the perspective of a corporate social media account aiming for maximum views, P12 noted that \textit{``they will give less views if they have the trigger warning'' (P12)} because warnings can reduce engagement and visibility.
This concern reflects the broader tension between maintaining content reach and protecting viewers from potentially distressing material.
P7 noticed that professional content creators often steer clear of certain words or content, including TW/CW, to avoid strikes or demonetization. 
In conclusion, the result highlights the perceivable trade-off that posters face between maximizing reach and fostering a considerate, informed viewing experience.

}

\revision{

\subsection{RQ3: How Can Social Media Platforms Enhance Users’ Experiences with Trigger and Content Warnings (TW/CW)?}
\label{4.3}
Social media platforms are also important stakeholders in the implementation of trigger warnings and content warnings because they set the structural foundation of warning mechanisms -- such as policy, primary modality, design, affordance, features, and algorithms. 
As previously seen, some platforms have built-in features that allow users to flag their content, which allows additional protection (a warning on top of a blurred image on X), while some platforms automatically detect sensitive content as part of content moderation (e.g., Instagram Nudity Protection).
Depending on the platform affordance and primary medium, users must adapt their practice of adding or discovering TW/CW.
In our study, participants suggested a number of changes they would like platforms to implement. In the following section, we present these suggestions in detail. 

}

\revision{
\subsubsection{\textbf{Desired Features for Viewers}}
During the interviews, participants discussed their desired features for TW/CW systems from the viewer's perspective based on their personal experiences.
Their needs include emotional support resources, warning system personalization, and warning explanations.

\textbf{Emotional Support Resources for Viewers Exposed to Warning-Labeled Content.} 
\label{doom-scrolling}
Throughout the interviews, participants expressed the need for additional resources to assist users when they are exposed to triggering content.
They highlighted that a behavioral response to consuming sensitive and triggering content, a tendency to fall into a cycle of seeking out more distressing information, in other words, doom-scrolling. 
Doom-scrolling is a state of media use typically characterized by individuals persistently scrolling through their social media newsfeeds with an obsessive focus on distressing, depressing, or otherwise negative information \cite{sharma2022dark}.
Such behaviors can emerge either when viewers begin with intentional or passive engagement.
As P14 described, \textit{``The idea of you want to know more, and it's giving feeding into that trigger a little bit.''}
While viewers may initially engage with warning-labeled content out of curiosity, this behavior can quickly lead to doom-scrolling, potentially influencing the algorithm to recommend even more similar content. 

To support viewers after exposure to triggering content, several participants (5/15) suggested that platforms should integrate with an SOS system, offering mental health resources and emergency contact information for crisis hotlines. 
This system could function similarly to the pop-up alerts for COVID-19 resources that were common on Facebook and Instagram during the pandemic. 
Such resources should be dynamic and responsive to current events or evolving users' personal needs.
As P14 explained, platforms should keep \textit{``updating them as needed''} to maintain relevance and effectiveness.


Despite the generalized support system, platforms can also provide a personalized emotional approach to self-care resources.  
For example, four participants mentioned that depending on the topic they are triggered by, they reach out to different people within their close circle for emotional support. 
Therefore, they proposed a feature that allows users to designate their trust contacts, such as friends and family members, as recommended support points when triggered by a social media post. 
P9 explained how peer support comforted them when they suffered from the triggering content.
\begin{quote}
    \textit{``When you talk to people and you understand that I have gone through the same process and all, it's not just me. Then you feel a little comforted.''} (P9)
\end{quote}    

Additionally, platforms could introduce features that recommend users to uplift their mood, such as creating personalized playlists or suggesting taking a break after detecting a series of heavy posts or videos.
For instance, five participants mentioned that watching comforting content or listening to music helps them calm down after being triggered or break the doom-scrolling cycle. 
P7 shared her experience; having a \textit{``disturbing stuff''}, such as a playlist on YouTube, helped her to stop. 
Participants also noted that platforms should proactively take responsibility for identifying the moments when viewers need a break.
Several participants (4/15) highlighted the importance for platforms to detect doom-scrolling behaviors and respond with interventions, such as a pop-up message or nudge (P7), when a user \textit{``might need to take mental breaks''} (P15).

Lastly, mindful practices such as \textit{``gratitude journaling''} (P15) and taking \textit{``decompression time''} (P10) to process after being triggered could also be incorporated into the platform’s self-care resources.
As P15 emphasized by adopting ``gratitude journaling'', she \textit{"understands people have really different experiences"} (P15).


\textbf{Personalized Content Moderation with TW/CW} 
\label{personalization}
Given the diversity of warning types (\ref{warning types}) and viewers' personal experience with triggering content(\ref{relevent to topics}), participants expressed their desires, from the viewers' perspective, for personalized warning systems to enhance their experience on social media platforms.
For instance, some participants (7/15) wished for more flexible and user-centered TW/CW mechanisms that would allow viewers to filter content based on their personal triggers.
This system allows viewers to select from a range of warning categories (e.g., abuse, self-harm, etc.) and customize which type of warnings they prefer to see or avoid.
Such a feature would not only protect users who are more vulnerable to triggering content but also improve the experience for those who are not easily triggered, as they could disable the setting and avoid repeatedly encountering unnecessary warnings.
As P8 emphasized, giving users control over their content experience is important, 
\begin{quote}
    \textit{``The best way to give the user control of what they see and what they choose to be filtered based on the input from other users who post.''} (P8)
\end{quote}
\noindent Since \textit{``everybody's unique and they have their own ways of dealing with things''} (P8), personalizing the resources would make them more effective. 

Furthermore, participants also discussed approaches to enhance the accessibility of TW/CW. 
For example, two participants suggested that video content should incorporate both text-based and spoken warnings, making individuals with disabilities (e.g., visually impaired, hearing impaired, or illiterate) more inclusive.

}

\revision{
\subsubsection{\textbf{Desired Features for Posters}}
Through our study, participants also highlighted several features from the posters' perspective that could better support and encourage users to engage with TW/CW systems.

\textbf{Proactive and Accessible TW/CW Features Enhancing Posters to Add Warnings for Triggering Content.}
Determining what content should be warning-labeled is an effort-consuming task for posters due to the variety of warning types.
Some participants (7/15) emphasized the importance of platforms taking a proactive role in encouraging posters to add warnings when sharing potentially sensitive content appropriately.
They explained that platforms should not allow sensitive content to be spread widely without a warning, and instead implement nudges that prompt posters to add warnings when necessary. 
P10 further suggested integrating such nudges into the user interface (UI) to streamline the process:
    \begin{quote}
        \textit{``If it was like a checkbox or something that asked you a question, maybe as you were making a post, leaving a comment. I think it'd be easier to not forget or like, have it like, built into the UI of whatever app you're using.''} (P10) 
    \end{quote}    
\noindent This approach reduces posters' mental effort and, in the meantime, helps minimize unintentional negligence.
By adopting this approach, social media platforms can consistently label triggering content.

Additionally, four participants suggested that the platform should give posters the tools to add specific categories of warnings, such as having a feature to add descriptive warning tags while making a post. 
Moreover, the addition of granularity settings on the poster's side for common triggers, categorizing them by natural causes and human fault, could be an efficient way for creates to add warnings.
P8 explained,
    \begin{quote}
        \textit{``Besides making it accessible and very easy, like when you go to post, maybe there is like a right on the side of the screen, just like the pull-down menu for like choosing them [the topic of warning].''} (P8)  
    \end{quote}

\textbf{Education and Awareness Campaigns Enhancing the Effectiveness of Using TW/CW.}
Participants suggested that the lack of awareness and education around warnings can be addressed by having discourses on their effective use. Recommendations included incorporating lessons on warnings into school curricula (P3), providing guidelines (P7), platform-led suggestions (P2, P9), and offering training (P15).

Participants were asked about having platform guidelines around warnings. According to P12, there should be a general basic guideline of what to avoid, but \textit{``people should be able to feel free to add more.''}
P7 wished for guidelines on approaching people in their circle to nudge them to add a warning.
However, it is challenging to establish universal or platform-wide guidelines and to capture people’s attention effectively.
On the one hand, developing such guidelines for warnings is difficult due to the nuanced nature of triggers and the diverse ways individuals interpret and respond to them.
Therefore, \textit{``there can be universal guidelines for certain triggers, but it will not be for everyone''} (P9) and \textit{"the choice should be in the user's hand''} (P15).
On the other hand, educational efforts to promote effective use of warnings struggle to capture users’ attention, as \textit{``no one reads it''} (P9) and liken them to terms and conditions. 
Instead, they tend to rely on instincts and familiar patterns from other applications.

Educating posters can be an effective strategy for platforms to help them better understand the concept of TW/CW, particularly for those encountering them for the first time.
However, users may sometimes accidentally be exposed to and suffer from triggering content.
Therefore, P2 thought that post-engagement metrics should be used by the platform to notice if \textit{``someone is posting triggering stuff''} (P2) to others and then send notifications to the author on \textit{how} to add a warning \textit{``using their tools.''}
Some platforms, such as Instagram or Facebook, have a history of promoting new features through stories and other in-platform advertising methods. 
P9 suggested employing a similar strategy to raise awareness about features to add TW/CW. 
This approach would facilitate user education without incurring significant costs for the platforms.


}

\section{Discussion}
\revision{


By reflecting on the findings, we proposed a framework (Figure~\ref{framework}) to better understand how the use of TW/CW can be improved on social media platforms. 
This framework captures the roles of three primary stakeholders: viewers, posters, and platforms.

Our findings revealed the challenges that viewers (\ref{4.1}) and posters (\ref{4.2}) encountered when they engaged with TW/CW.
We also found participants' needs (\ref{4.3}) for the TW/CW mechanisms, highlighting the opportunities for social media platforms to support their needs.
Social media platforms are uniquely positioned in this mechanism as they are not only the carriers but also a bridge between the posters and viewers.
As such, we aimed to explore the relationships between the platforms and posters and viewers. 
For example, the challenges faced by posters and viewers are tightly connected through the limited affordance of a social media platform and its variance across platforms, as explored more in the previous sections (\ref{4.1}, \ref{4.2}).
In this section, we build upon these insights to discuss how social media platforms can better support the TW/CW mechanisms.

\label{Figures}
\begin{figure}[t]
  \centering
  \includegraphics[width=\textwidth]{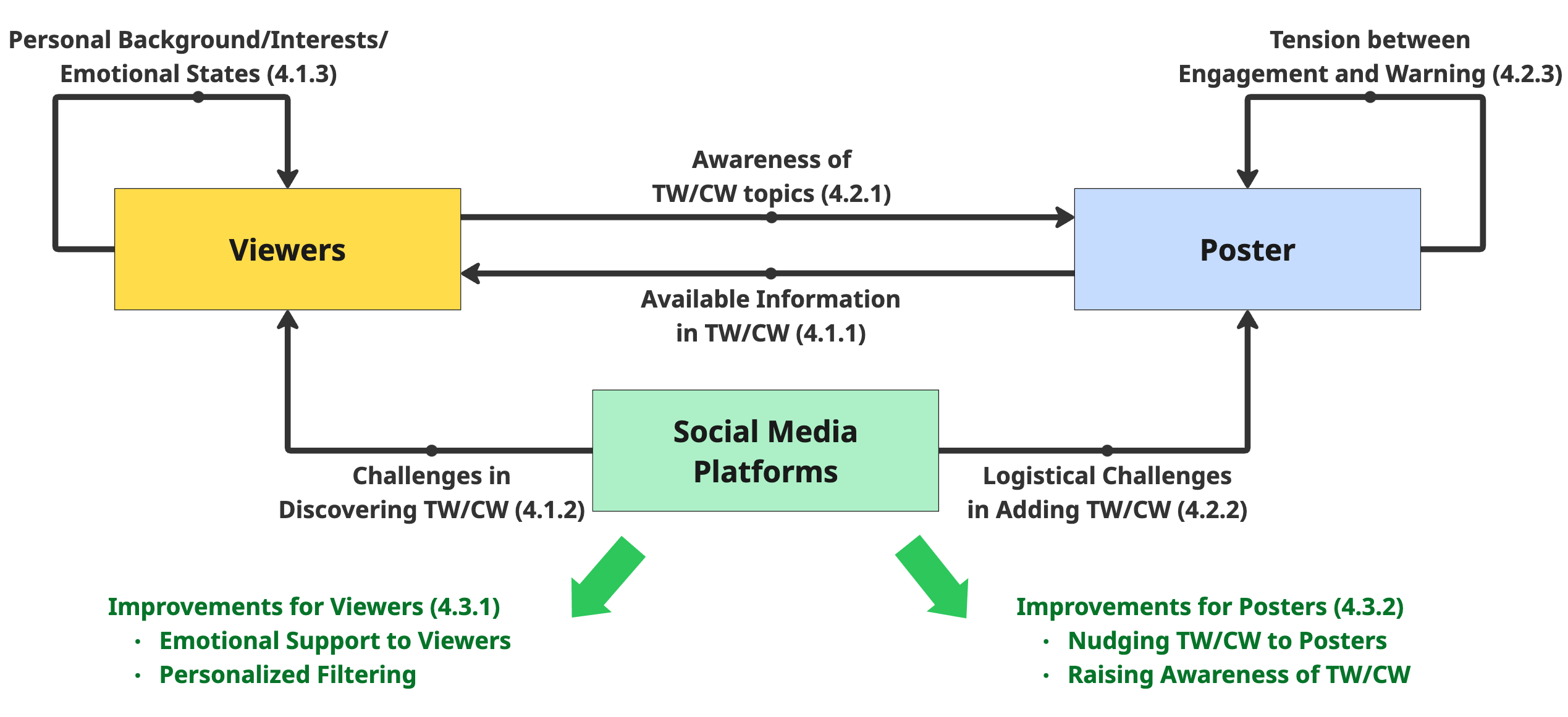}
  \caption{Framework of the Trigger Warning (TW) and Content Warning (CW) Mechanism}
  \label{framework}
\end{figure}
}
\revision{
\subsection{Integrating Social Media Posters into the Content Moderation Ecosystem through TW/CW}

Our research introduces a novel perspective for online safety by shifting focus away from traditional content moderation approaches that primarily rely on platforms and viewers. Existing methods include platform-centric moderation, where platforms implement automated systems to detect harmful content that needs to be moderated~\cite{horta2023automated, gorwa2020algorithmic}; community-based moderation, relying on community members or bots to monitor and regulate content~\cite{jhaver2023bans, tran2024challenges, geiger2016bot, jhaver2019human}; and personalized moderation for viewers~\cite{jhaver2019does, jhaver2023personalizing}. In contrast, our study advances the understanding of how posters or content creators can take a proactive role in contributing to content moderation efforts, leveraging their awareness and willingness to protect their audience from distressing content.

Our findings suggest that users are willing and desire to protect other social media users from distressing and triggering content despite the additional efforts needed to add content warnings~\cite {haimson2020trans}. 
In addition, users are also willing to take the perceivable risk of being curated out by machine learning algorithms~\cite{gillespie2022not} or personalized filters~\cite{jhaver2018online}.
Our result is consistent with the previous work that showed the tension between content moderation, free speech, and the suppression of marginalized groups --- such as trans and Black social media users~\cite{seering2020reconsidering, jhaver2023personalizing, haimson2021disproportionate}.
In turn, using TW/CW can potentially provide a security measure to posters as well, mitigating greater consequences such as violating a community policy or being blocked by other users. 
Still, on some platforms, features equivalent to TW/CW (e.g., Sensitive Content in Instagram or TikTok) are implemented through platform-centric, automated detection algorithms without providing authors an opportunity to flag their content as sensitive content. The limitation undermines the posters' agency, which encourages and supports proactive participation in the moderation process using TW/CW. 

However, it is worth noting that TW/CW mechanisms are limited in scope, as they apply the cases in which the poster’s intent is not malicious. For example, toxic behaviors, hate speech, or online harassment are typically addressed through community-based or machine learning–based moderation approaches~\cite{chandrasekharan2017you, jhaver2018online}, as we cannot expect individuals exhibiting such behavior to use TW/CW voluntarily. In these cases, the content creator’s harmful intent precludes the use of self-reported warnings.


Still, we anticipate that nurturing the norm of using TW/CW can positively impact existing content moderation approaches in multiple ways. 
Content warnings can become a community norm, which may alleviate the burden of community-based content moderation and foster a more thoughtful online culture~\cite{haimson2020trans}.
Furthermore, raising awareness of distressing content can also enhance individuals’ capacity to moderate other posters’ content. 
From the data-centric perspective, TW/CW can improve machine learning–based detection approaches by leveraging self-reported labels provided by posters, similar to personalized filtering words~\cite{10.1145/3491102.3517505}. 
Unlike platform-directed moderation or personalized filtering tools that may provoke concerns about censorship or diverging user preferences, TW/CW offers a self-imposed, author-driven approach that avoids such tensions. Prior work has shown that users weigh free speech concerns and perceived harms to others when evaluating moderation systems~\cite{jhaver2023personalizing}. In contrast, TW/CW relies on voluntary disclosures by posters and is less likely to be viewed as an infringement on others’ expression or autonomy.
Overall, using TW/CW offers additional protection for individuals from exposure to distressing content, and in doing so, TW/CW contributes to a more holistic content moderation ecosystem -- one that balances authorial intent, viewer protection, and platform minimal intervention.
To that end, our results contribute to supporting more effective TW/CW practices by identifying key challenges and areas for improvement. 

}

\revision{
\subsection{Protecting Viewers from TW/CW Contextual Information}
The level of specificity in TW/CW labels can significantly affect viewers' emotional responses, underscoring the need for more thoughtful design practices. Participants expressed concern that revealing overly specific details—whether through keywords or blurred images that hint at the underlying content—can itself be distressing. Previous research has advocated for more specific warnings to capture the nuances of a post~\cite{haimson2020trans}. ~\citeauthor{south2023exploratory} proposed that providing warnings at multiple levels of detail can support viewers’ decision-making, particularly in the context of photosensitive content~\cite{south2023exploratory}.  

These works collectively suggest the value of multi-level content warnings, one general and one more specific, to offer layered contextual information that empowers viewers to decide how much detail they want to access. 
Our findings extend the current research by suggesting that overly detailed warnings may have unintended, potentially triggering effects.
Individuals who have experienced traumatic events and are in a phase of trauma denial may find that encountering a content warning inadvertently retraumatizes them~\cite{church2017childhood}. In such cases, providing highly specific information—through detailed keywords or partially obscured images—can intensify this effect by making the underlying content more apparent and triggering. Therefore, while generic content warnings are intended to protect viewers, their implementation requires careful consideration to avoid unintended harm.
One design implication can be that the contextual information available along with TW/CW could be guided by the platform and be presented optionally in response to a viewer’s interaction. For instance, the ``See Why” button in Figure~\ref{slide3}-B exemplifies this approach, where a general label is initially displayed, and viewers have the option to access a more detailed explanation if they click the button.
}

\revision{
\subsection{Rethinking the Role and Design of Trigger and Content Warnings for Viewers}
According to our results, TW/CW labels might provoke curiosity and increase engagement with distressing content, thus defeating their purpose.
Multiple studies have found that TW/CW labels do not effectively mitigate emotional responses or help users avoid distressing content and may instead increase anticipatory distress~\cite{bridgland2022meta}. In another study, Instagram’s blurred screen feature failed to regulate users’ desire to view distressing content in an experimental setting, regardless of their vulnerability~\cite{bridgland2023curiosity}. 
Even more concerning, research has shown that individuals with traumatic experiences sometimes engage in self-triggering behaviors, actively seeking out disturbing content online~\cite{bellet2020self} or exhibit patterns of doom-scrolling~\cite{ sharma2022dark}. In such cases, TW/CW labels may unintentionally serve as an easier pathway for locating that content.
This tendency contradicts the traditional notion of social translucence~\cite{erickson2000social} in which increased awareness of potential harm does not necessarily translate to harm reduction and might even exacerbate it for certain individuals.  
These findings, confirmed by experiential studies, raise important questions about the actual protective value of the current TW/CW practice and whether current implementations genuinely serve the needs of those they aim to protect, or inadvertently contribute to further harm.

Despite the questionable effectiveness of TW/CW, such information can still be valuable as part of a broader system of online safety. One new way to leverage TW/CW labels is to provide crucial metadata that allows viewers to block specific types of content entirely, depending on their current emotional state or preferences~\cite{jhaver2023personalizing}.
The existing scenario involves TW/CW being used as part of the content presentation, typically in a social media feed,  and the viewer chooses whether to engage with it. TW/CW labels and the topic behind the warning can be integrated into a personalized moderation mechanism.
Such blocking features can benefit individuals who are susceptible to distressing content, much like distraction-blocking applications support users in productivity contexts~\cite{mark2018effects}.

Participants also expressed a desire for personalized content filtering based on their individual triggers (see~\ref{personalization}), aligning with the recommendations of the trauma-informed computing framework~\cite{chen2022trauma}. For instance, \citeauthor{heung2025ignorance} explored how disabled users prefer configuring filters to manage exposure to ableist speech, highlighting the importance of personalized moderation tools in mitigating harm on social media platforms~\cite{heung2025ignorance}. This approach can be especially valuable in cases where lay social media users might not anticipate that certain content could be perceived as distressing by others. 
Another common example involves individuals with substance use disorders managing digital triggers that may appear harmless to most people, such as a social media post of someone drinking at a bar~\cite{phelan2022work}.
Linking TW/CW labels with personalized filtering systems would allow users to automatically exclude specific types of content from their feed, eliminating the need to make decisions on a post-by-post basis.

}

\revision{
\subsection{Supporting Posters' Effective Use of TW/CW}

Our study revealed that social media posters face several challenges in effectively adding TW/CW to their content. Most notably, they may lack a clear or comprehensive understanding of which topics warrant such warnings. Additionally, logistical difficulties can arise due to the wide variety of methods available, making it difficult to apply TW/CW consistently. 
The consequences of ineffective or missing TW/CW can be significant: content may be blocked by other users, flagged by the platform, and--most importantly--may unintentionally trigger traumatic memories of others.
In this regard, we find a knowledge gap in supporting posters' effective use of TW/CW through behavioral interventions and design changes in social media platforms. 

Raising social media users' awareness of content has been extensively studied, particularly in the context of news consumption. Research has focused on identifying misinformation, mitigating its spread, and enhancing users' ability to assess the credibility of online content~\cite{bhuiyan2021nudgecred, 10.1145/3613904.3642490, jahanbakhsh2021exploring, jahanbakhsh2023exploring}. While most interventions target users during content consumption, there is an opportunity to educate content creators before they post. Pre-submission feedback mechanisms, common in writing tools like Grammarly, can be adapted to social media platforms. Implementing just-in-time educational prompts could guide users on whether and how to apply trigger or content warnings (TW/CW) to their posts. For example, a large language model (LLM) can analyze a draft post to determine if it pertains to sensitive topics outlined in established typologies~\cite{charles2022typology}. If so, the system could suggest or even auto-complete a TW/CW label with specific information (e.g., ``TW/CW: This post contains a story of domestic abuse''). Such proactive interventions offer a novel approach to mitigating risks, moving beyond traditional post hoc moderation methods and empowering content creators through opportunistic learning.

Another improvement for content creators, as suggested by our study results, is facilitating the addition of TW/CW beyond textual labels, especially when the primary modality of social media is not text. While some platforms allow users to self-select sensitive content options that automatically blur their images, manually incorporating TW/CW into visual content—such as adding a few seconds of video with TW/CW labels and voiceover for YouTube videos or creating a separate first slide for slideshow content—requires additional effort. Platforms can support the integration of TW/CW as part of visual content by enabling features that automatically generate content warnings before the main visual content. Although platforms may automatically add such visuals upon detecting explicit content, users currently lack the ability to voluntarily add such warnings for their viewers. This approach empowers posters to proactively safeguard their audience. 
}

\revision{
\subsection{Beyond Warnings: Emotional Support Systems for Post-Exposure Care}
While the literature has predominantly focused on the effectiveness of warnings, our research has highlighted the limitations of TW/CW and the challenges of protecting viewers through TW/CW due to the complexity involved in viewers' decisions. 
Therefore, social media platforms should complement the limitation of TW/CW by prioritizing strategies aimed at assisting users \textit{after} they have encountered distressing content. This perspective aligns with findings by \citeauthor{phelan2022work}, which suggest that reducing sensitivity to triggers is crucial for recovery, especially since exposure to distressing content is inevitable~\cite{phelan2022work}. \citeauthor{scott2023trauma} suggested trauma-informed social media and provided design choices that social media can make to track and respond to viewers' harmful online experiences~\cite{scott2023trauma}.
Similar to our participants' suggestions (music playlist, comforting videos, etc), we propose that platforms introduce mindful features to help users build resilience and cope with the overwhelming feelings that can follow a triggering experience. 

The social media platform can offer interventions and services that support mental health, drawing from clinical psychology and other HCI works that intervene users for their mental health.  
Integrating therapeutic techniques, such as cognitive-behavioral therapy (CBT), into digital platforms has shown promise in the context of HCI to mitigate mental health of various domain, including patient care~\cite{doherty2012engagement}, higher-education~\cite{fitzpatrick2017delivering}, knowledge work, ~\cite{howe2022design, chow2023feeling}, eating disorders~\cite{devakumar2021review}. 
For example, following exposure to content with TW/CW, platforms could offer users access to brief, evidence-based digital CBT intervention, providing immediate relief with the context and empowering users to build resilience over time.

Other HCI work focusing on facilitating peer-support systems or nudging users to engage with existing mental health resources can be applied to provide aftermath support for those who experience distress after viewing sensitive content. \citeauthor{cohen2023improving} investigated the impact of embedding brief, enhanced crisis response interventions within social media platforms, demonstrating increased access to mental health resources compared to standard practices that merely provided a crisis hotline link~\cite{cohen2023improving}. \citeauthor{mahar2018squadbox} presents another relevant example, in which an interactive system mediates peer support by coordinating trusted individuals to assist users facing online harassment via email~\cite{mahar2018squadbox}. In the social media context, \citeauthor{patton2018accommodating} examined how gang-involved youth use Twitter to express grief and mourning after the death of close friends or family, highlighting the complex ways in which social media serves as a space for emotional expression and community support among marginalized populations~\cite{patton2018accommodating}. However, to apply peer-support systems to those who suffer from distressing content, sharing private data (e.g., viewing history) within a social context may limit the feasibility of facilitating peer support with close others, due to concerns about confidentiality and stigma~\cite{andalibi2020disclosure, patton2018accommodating}.

Overall, social media platforms can complement the limited effectiveness of TW/CW systems by designing features that support users after exposure to triggering content, potentially in conjunction with the use of warnings.
}

\subsection{Limitations and Future Work} \label{sub:limitations}
\revision{
One limitation of our study pertains to participant selection. The majority of participants we recruited exhibited a pre-existing interest in, and a proactive approach toward, the use of trigger/content warnings (TW/CW). Consequently, we faced challenges in recruiting individuals with opposing viewpoints—those who oppose the use of TW/CW or who have never engaged with or reflected on them. Additionally, our findings are based on casual social media users and do not capture the perspectives of individuals who have experienced severe trauma, as we did not include such eligibility criteria (e.g., veterans or survivors of sexual assault) and did not inquire if they have. 

Furthermore, the participants we recruited were not necessarily heavy content creators (e.g., YouTubers, TikTokers, or influencers), whose perspectives may differ significantly from those of casual users. Finally, existing literature has largely overlooked content creators' perspectives regarding the use of TW/CW. Future research should delve deeper into testing strategies to address these challenges and prioritize the prototyping and user testing of nudging and intervention techniques aimed at encouraging the addition or viewing of TW/CW. Such studies will help determine the efficacy and impact of these methods, ultimately fostering a more inclusive and supportive online environment.

}

\section{Conclusions}

\revision{Our study provides an in-depth exploration of social media users' perceptions regarding the use of trigger warnings (TW) and content warnings (CW) in the social media landscape. 
Through 15 interviews, we delved into viewers' and posters' challenges in employing TW/CW, the factors influencing their decision-making processes, and design recommendations to enhance users' experiences with warnings. 
We further identify the unique role that social media platforms can play in the TW/CW ecosystem.
Based on the findings, we propose a conceptual framework of TW/CW mechanisms, which demonstrates the relationships among viewers, posters, and social media platforms. 
The framework also shows how social media platforms can better support viewers' and posters' challenges and streamline their experiences when interacting with TW/CW. 
Finally, we offer design implications for social media platforms to support posters' effective use of TW/CW and to protect viewers' mental health.


}


\bibliographystyle{ACM-Reference-Format}
\bibliography{references}


\appendix

	
\section{Appendix}  

\subsection{\textbf{Study Survey}}
\label{sec:survey}

\noindent{\textit{Screening Questions}}
\begin{itemize}
    \item Are you over the age of 18? (Choose one.)
    \begin{enumerate}
        \item Yes
        \item No
    \end{enumerate}

    \item Are you currently a US citizen or resident (meaning legally authorized to live in the US, like being on F1 student visa or H1B)? 

    (Choose one.)
    \begin{enumerate}
        \item Yes
        \item No
    \end{enumerate}

    \item I consent to my involvement in this survey. (Choose one.)
    \begin{enumerate}
        \item Yes, I consent to my involvement in this survey
        \item No, I want to quit the survey.
    \end{enumerate}
\end{itemize}

\noindent{\textit{Demographic Questions}}
\begin{itemize}
    \item Age
    \item Race/Ethnicity (Check all that apply.)
    \begin{enumerate}
        \item Asian or Pacific Islander
        \item Black or African American
        \item Hispanic or Latino
        \item Native American or Alaskan Native
        \item White or Caucasian
        \item Other
    \end{enumerate}

    \item How do you describe your gender? (Choose one.)
    \begin{enumerate}
        \item Woman/Female
        \item Man/Male
        \item Non-binary
        \item Prefer not to answer
        \item Other
    \end{enumerate}

    \item Are you transgender? (Choose one.)
    \begin{enumerate}
        \item Yes
        \item No
        \item Prefer not to answer
    \end{enumerate}
\end{itemize}

\noindent{\textit{Your Experience With Trigger/Content Warnings}}
\begin{itemize}
    \item What social media platforms do you regularly use? (For purposes of this study, we are not considering strictly messaging sites/apps, like iMessage or Facebook Messenger.)

    (Check all that apply)
    \begin{enumerate}
        \item Facebook
        \item Instagram
        \item Snapchat
        \item Reddit
        \item TikTok
        \item Discord
        \item Pinterest
        \item Twitter
        \item YouTube
        \item Tumblr
        \item Medium
        \item LinkedIn
        \item Quora
        \item Twitch
        \item VSCO
        \item BeReal
        \item Imgur
    \end{enumerate}

    \item How often do you view content on social media? (Choose one.)
    \begin{enumerate}
        \item Never
        \item Rarely
        \item Sometimes
        \item Often
    \end{enumerate}

    \item How often do you see trigger and/or content warnings on others' social media posts? (Choose one.)
    \begin{enumerate}
        \item Never
        \item Rarely
        \item Sometimes
        \item Often
        \item Always
    \end{enumerate}

    \item How often do you post on social media? (Choose one.)
    \begin{enumerate}
        \item Never
        \item Rarely
        \item Sometimes
        \item Often
    \end{enumerate}

    \item How often do you add trigger and/or content warnings to your social media posts? (Choose one.)
    \begin{enumerate}
        \item Never
        \item Rarely
        \item Sometimes
        \item Often
        \item Always
    \end{enumerate}

    \item Are you interested in being interviewed to share more about your answers? * The interview would be held on Zoom for 1-1.5 hours and you would be compensated with a \$15 Amazon gift card. (Choose one.)
    \begin{enumerate}
        \item Yes
        \item No
    \end{enumerate}
\end{itemize}

\noindent{\textit{Interest in Being Interviews}}
\begin{itemize}
    \item Name
    \item Email
    \item Since this research deals with trauma and triggers, sensitive subjects may arise. Are there any topics you do not wish to discuss at all during the interview?
\end{itemize}

\revision{
\subsection{\textbf{Initial Interview Protocol for P1-P6}}
\label{sec: initial_interview}

\noindent\textit {Participant’s use of social media}
\begin{itemize}
    \item What social media platforms do you most commonly use?  
    \item What do you typically post on <platform>?  What do you typically view on <platform>?  
    \item Who is your audience on <platform>?  Who do you typically follow on <platform>?  
        \begin{enumerate}
            \item On <platform>, is your account public or private?
            \item On <platform>, do you typically post to a universal audience or to individual sub-communities/groups?
            \item On <platform>, do your friends/followers know who you are, or are you anonymous?
        \end{enumerate}
    \item When you post on <platform>, do you post on a wide variety of topics or in specified groups/pages with more narrow purposes?  When you view content on <platform>, are you viewing specific topics or pages with a wide variety of topics?  
\end{itemize}

\noindent{\textit {Social media user as a viewer/consumer with current UI}}

\begin{itemize}
    \item First, we will be talking about what you see as you spend time on different social media platforms.  Do you often see trigger or content warnings on social media?

    \item If/when you do, what do these warnings typically look like?  What does it say?  Where is it located?

    \item What topics usually have a trigger or content warning present?  When would you expect to see either a trigger or content warning present on a post?

    \item Have you noticed any difference between someone adding a “content warning” and a “trigger warning”?  Do they cover different topics?  Do you see one term more frequently?  Are the terms “trigger warning” and “content warning” used interchangeably?

    \item Is the difference between trigger warnings and content warnings clear to you?  Is it even important for there to be a distinction between them?
        \begin{enumerate}
            \item What topics usually have a trigger warning present?  What topics usually have a content warning present?  When would you expect to see a trigger warning present on a post? When would you expect to see a content warning present on a post?
        \end{enumerate}

    \item How do trigger and content warnings change your viewing/reading behaviors on social media?  Do you primarily ignore them and read anyway, do you not view/read posts that could be triggering to you, do you save for later when your mental health may be in a better place or where you are in a safe or private space?

    \item Have you ever read a post that had a trigger warning or content warning present and you chose to read it anyway and the content was unexpected? Why was it unexpected?  How did you react?
        \begin{enumerate}
            \item Have you ever seen a TW/CW that didn't match the content it was supposed to be about?
            \item Have you ever seen a TW/CW and decided to see the post anyway (curiosity maybe) and you underestimated the effect it had on you?
        \end{enumerate}

    \item Have you ever read/viewed something on social media that you wish had a trigger warning or content warning present?

    \item Are there any topics you find to be personally triggering or you wish to avoid that are often not covered by trigger or content warnings?
\end{itemize}

\noindent{\textit{Social media user as a poster/creator with current UI}}

\begin{itemize}
    \item Now, we are going to be discussing how users can add their own warnings to posts.  Do you often add trigger or content warnings to your posts?  How do you decide when to use those?

    \item If/when you do add trigger or content warnings to your posts, what do these warnings typically look like?  What does it say?  Where is it located? 

    \item For what topics, will you usually add a trigger or content warning?

    \item Do you add both “trigger warnings” and “content warnings”?  Is there any difference between them to you?  Do they cover different topics?  Do you use one term more frequently?  Are the terms “trigger warning” and “content warning” used interchangeably in your posts?

    \item Is the difference between “trigger warnings” and “content warnings” clear to you?  Is it even important for there to be a distinction between them?
        \begin{enumerate}
            \item For what topics, do you usually add a trigger warning?  For what topics, do you usually add a content warning?
        \end{enumerate}

    \item How does adding trigger and content warnings change your posting style on social media?  Does it not affect it at all?  Does it make you feel more comfortable when posting about sensitive topics that everyone may not want to see?

    \item Is posting helpful trigger or content warnings when needed something that is important to you? 

    \item Is there any hesitation or difficulty when deciding whether to add trigger or content warnings to your posts?  If so, what are they? 

    \item Were there any posts that you later realized that you should’ve added trigger or content warnings?  How did you realize this?
    
    \item Do you have a clear understanding on what kinds of topics should have trigger or content warnings?

\end{itemize}

\noindent{\textit{Social media user and their ideal UI}}

\begin{itemize}

    \item Now, we are going to be discussing the ideal way you would like to see trigger and content warnings shown on social media. How do you wish trigger warnings and content warnings were shown on social media?  What would they say?  What would they look like?  Where would they be located in a post?

    \item Are there any ways that other social media users could better add warnings to be more considerate to you and your personal triggers?

    \item Would universal guidelines, or at least platform-wide guidelines, on how and when to add trigger or content warnings benefit you as a social media user?

    \item Trigger warnings and content warnings are a preventative measure, if you do get triggered then what practices do you follow after that to calm down? Can they be replicated digitally?
\end{itemize}

\noindent{\textit{Wrap-Up}}

\begin{itemize}
    \item Is there anything we didn’t talk about on the subject of trigger and content warnings on social media that you would like to share?
\end{itemize}
}

\subsection{\textbf{Updated Interview Protocol for P7-P15}}
\label{sec: interview}

\noindent\textit {Participant’s use of social media}
\begin{itemize}

\item Just for background, what is your use of social media like? What sort of apps, what content do you view, and who is your audience?  

 \end{itemize}

\noindent{\textit {Social media user as a viewer/consumer with current UI}}

\begin{itemize}
    \item First, we will discuss what you see as you spend time on social media platforms. What is your experience with seeing TW/CW on social media?

    \item If/when you do see them, what do these warnings typically look like?  What do they say?  Where are they located? Can you give an example? 
    
    \revision{
    \textbf{(Show warning examples to the participants)}
        \begin{enumerate}
            \item What scenarios out of these have you seen on your social media usage?
            \item Which method do you find the most effective as a viewer? Why?
            \item Which way will you use as a poster? Again, why?
            \item What can go wrong in such scenarios?
        \end{enumerate}
}

    \item What topics usually have a TW/CW present?  Describe the scenarios for expecting a TW/CW present on a post.

    \item How do TW/CWs change your viewing/reading behaviors on social media? Do you primarily ignore them and read them anyway? Do you not view/read posts that could be triggering to you? Do you save for later when your mental health may be in a better place or where you are in a safe or private space? Why?

    \item How do you decide to view or skip a post with TW/CW? Do you want them as a viewer? Why or why not?

    \item What has been your experience seeing a post with TW/CW and deciding to view the content anyway?

        \begin{enumerate}
            \item Has the content ever been unexpected behind the warning? Why was it unexpected? How did you react?
            \item The content did not match the warning
            \item Underestimating the effect it had on you
        \end{enumerate}

    \item What has your experience been on reading/viewing something on social media that you wish had a TW/CW present? Any specific topics?

    \item How do you currently avoid triggering content for you on social media?
\end{itemize}

\noindent{\textit{Social media user as a poster/creator with current UI}}

\begin{itemize}
    \item Now, we will discuss how you, as a user (can) add warnings to posts. What is your experience with using TW/CW on social media?  How do you decide when to use those?  \textit{(Most likely, it’s relevant to their answers from 1 so refer to those answers.)}

    \item If/when you do add TW/CW in your posts, what do these warnings typically look like?  What does it say?  Where is it located? Can you give an example? \textit{(This could be different depending on the type of post - image with caption, video, text-post - and possibly even the platform.)}

    \item For what topics do you use a TW/CW? Describe the scenarios when you expect to add them to your post/comment. What are the topics that you are fine with but still use TW/CW?

    \item How does using TW/CW affect your posting style on social media? Does it not affect it at all? Does it make you feel more comfortable when posting about sensitive topics that some folks might be triggered by?

    \item Is posting TW/CW helpful or needed when something is important to you?

    \item What are your views on putting TW/CW when posting content (including comments) on social media? Why do (or don’t) you use them?

    \item What are your thoughts on experiencing hesitation or difficulty when deciding whether or how to add TW/CW to your posts? Why do you think it is there? 

    \item What’s your experience on realizing later on that you should’ve added a TW/CW to your post?  How did you realize this? If not, why do you think that is? 

    \item What kinds of topics should have TW/CWs?

    \item What are your thoughts on the difference between viewing or using a “content warning” and a “trigger warning”? Do they cover the same or different topics? Are the terms used interchangeably? Do you see or use one term more frequently? Which one? (If yes, give examples - both viewer and poster perspective)

\end{itemize}

\noindent{\textit{Social media user and their ideal UI}}

\begin{enumerate}
    \item Short version
    \begin{itemize}
        \item Now, we are going to be discussing the ideal way you would like to see TW/CW shown on social media. What do you think the platform should do? Should it do anything? or should it leave it completely to users? If so, why or why not?
    \end{itemize}

    \textbf{OR}

    \item Longer version if time permits
    \begin{itemize}
        \item Now, we are going to be discussing the ideal way you would like to see TW/CW shown on social media. What are your frustrations so far with TW/CW on social media?

        \item How do you wish TW/CWs were shown on social media?  What would they say?  What would they look like?  Where would they be located in a post?

        \item What better ways can other social media users add warnings to be more considerate of you and your triggers?

        \item What are your views on having universal guidelines, or at least platform-wide guidelines on how and when to add or view TW/CW? Will they be helpful or not useful? Why?

        \item TW/CW are preventative measures; how about aftermath measures? (practices to calm down)
    \end{itemize}
 
\end{enumerate}

\noindent{\textit{Wrap-Up}}

\begin{itemize}
    \item Is there anything we didn’t talk about on the subject of TW/CW on social media that you would like to share?
\end{itemize}

\subsection{\textbf{Warning Examples}}
The following are TW/CW examples shown to the participants in interviews.
\label{slideshow}

\begin{figure}[H]
        \centering
        \includegraphics[width=\textwidth]{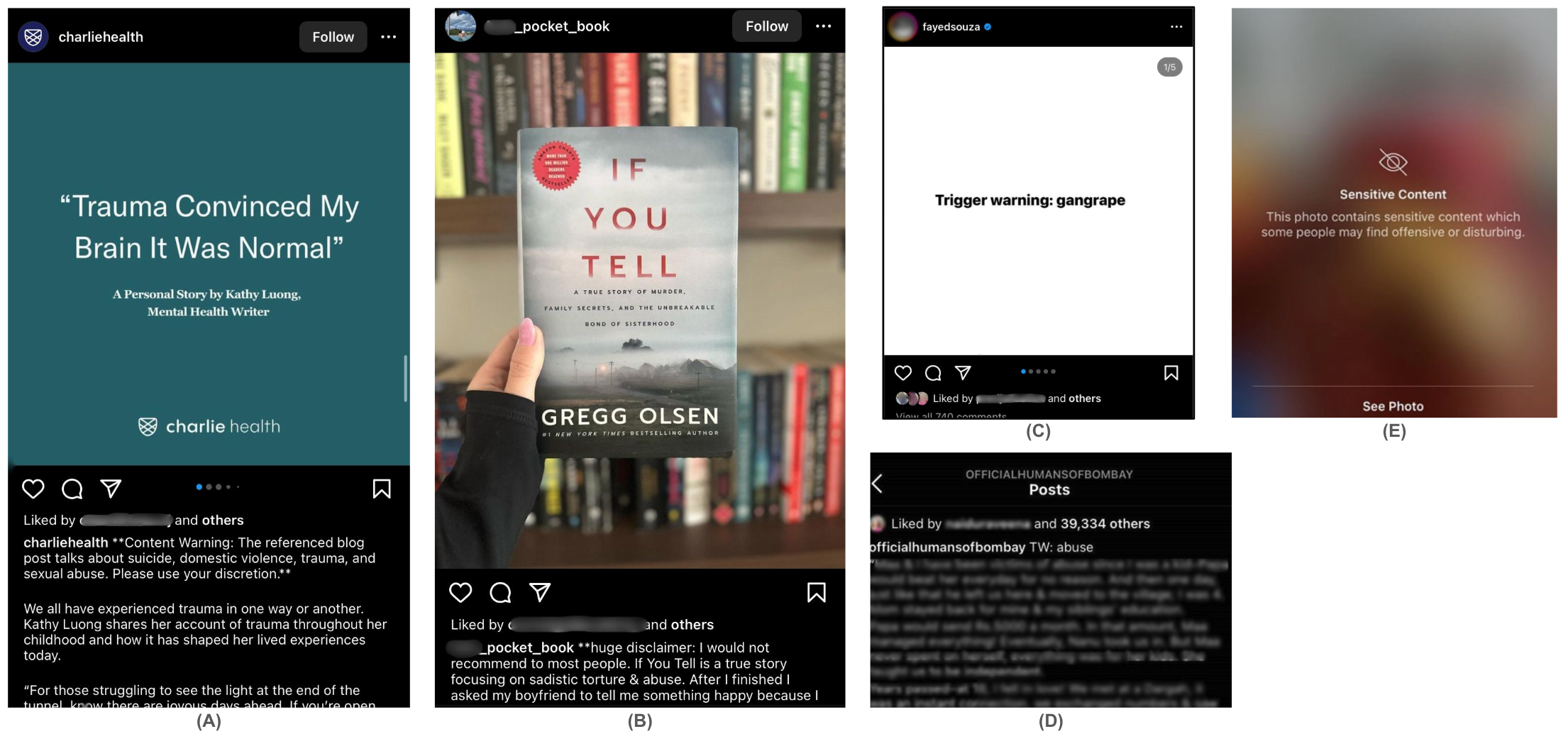}
        \caption{(A) CW in caption text post on Instagram, (B) Informal CW in caption text post on Instagram, (C) TW added in the first slide of post on Instagram, (D) TW embedded within caption text on Instagram, (E) CW for a post on Instagram}
        \label{slide1}
\end{figure}

\begin{figure} [h]
        \centering
        \includegraphics[width=0.9\textwidth]{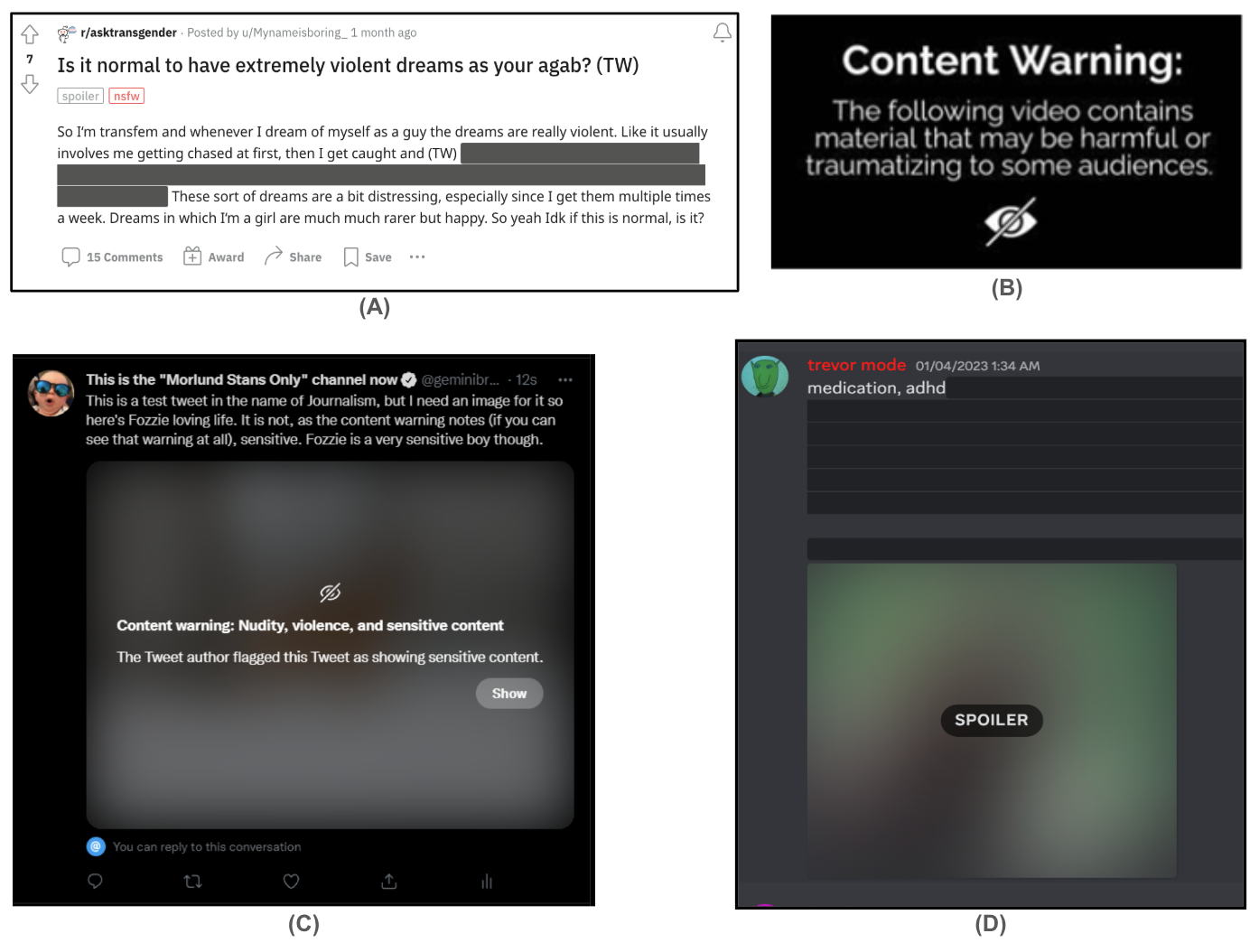}
        \caption{(A) CW embedded within text post on Reddit, (B) CW for a video on Facebook, (C) CW for a post on Twitter, (D) CW for a post on Discord}
\label{slide2}
\end{figure}

\begin{figure} [h]
        \centering
        \includegraphics[width=\textwidth]{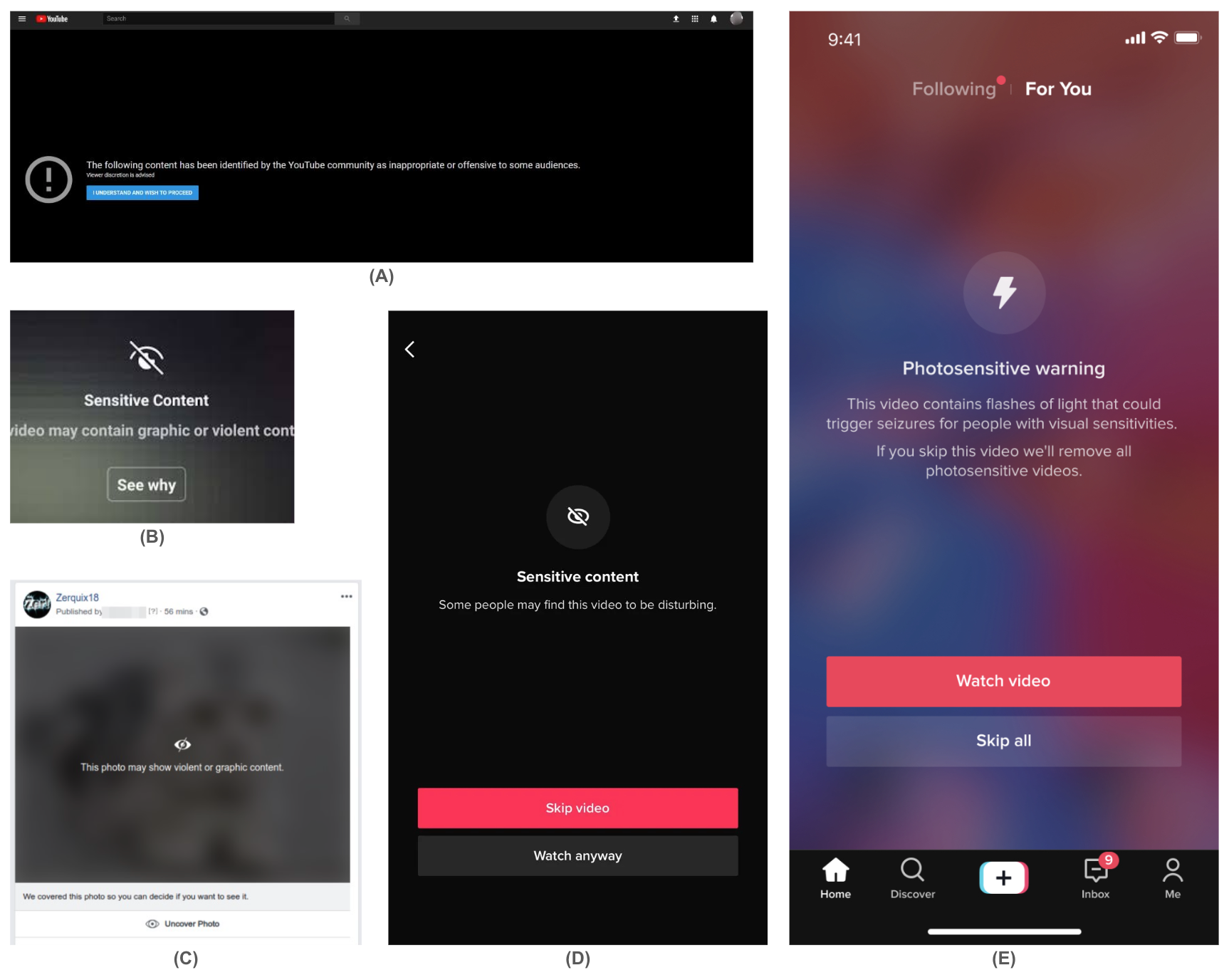}
        \caption{(A) CW for a video on YouTube, presented with a single-option interface, (B) CW for a video on Facebook, presented with an explanation option, (C) CW for an image on Facebook, presented with a single-option interface, (D) CW for a video on Facebook, presented with a dual-option interface, (E) CW for a video on TikTok, presented with a dual-option interface}
\label{slide3}
\end{figure}

\revision{
\subsection{\textbf{Participant Background on Social Media Usage}}
\label{background}
\vspace{10pt}

\begin{table} [h]
\setlength{\tabcolsep}{8pt}
\resizebox{\textwidth}{!}{%
\begin{tabular}{c p{4cm} p{1.6cm} p{2.5cm} p{1.6cm} p{2.5cm}} 
\toprule
\textbf{Participant ID} & \textbf{Social Media Platforms Used} & \textbf{Posting Frequency} & \textbf{TW/CW Posting Frequency} & \textbf{Viewing Frequency} & \textbf{TW/CW Viewing Frequency}\\ [1ex] \midrule
P01 & Facebook, Discord, Twitter, YouTube, Tumblr, Imgur & Sometimes & Sometimes & Often & Often\\ 
P02 & Facebook, Instagram, Reddit, TikTok, Pinterest, LinkedIn, BeReal & Sometimes & Never & Often	& Often\\ 
P03 & Instagram, Twitter, YouTube, LinkedIn	& Often	& Often	& Often	& Often\\ 
P04 & Facebook, Instagram, Reddit, Twitter, YouTube, LinkedIn, Twitch & Sometimes & Never & Often & Often\\
P05 & Facebook, Instagram, Reddit, Discord, Twitter, YouTube, LinkedIn, Quora, Twitch & Sometimes & Rarely	& Often	& Often\\
P06 & Facebook, Instagram, Snapchat, Reddit, TikTok, Discord, Twitter, YouTube, LinkedIn, Twitch & Often & Always	& Often	& Always\\
P07 & Facebook, Discord, YouTube, LinkedIn	& Often	& Always & Often & Sometimes\\
P08 & Facebook, Snapchat, Reddit, Discord, Twitter, YouTube, LinkedIn, Twitch & Often & Sometimes & Often & Sometimes\\
P09 & Facebook, Instagram, Snapchat, Discord, Twitter, YouTube, Medium, LinkedIn & Sometimes & Never & Often & Always\\
P10 & TikTok, Discord, Twitter, YouTube, Twitch	& Never	& Never	& Often	& Sometimes\\
P11 & Facebook, Instagram, Snapchat, Reddit, TikTok, Discord, Pinterest, Twitter, YouTube, LinkedIn & Often & Often & Often & Sometimes\\
P12 & Facebook, Instagram, Reddit, YouTube, LinkedIn & Rarely & Never & Rarely	& Sometimes\\
P13 & Instagram, YouTube, LinkedIn, Quora & Sometimes & Never & Often & Sometimes\\
P14 & Instagram, Snapchat, Discord, Twitter, YouTube, LinkedIn	& Often	& Never	& Often	& Sometimes\\
P15 & Facebook, Reddit, Twitter, YouTube, Medium, LinkedIn, Quora & Sometimes & Sometimes & Often	& Sometimes\\
\bottomrule
\end{tabular}}
\caption{Participant Background on Social Media Usage}
\label{table:3}
\end{table}
}


\end{document}